\documentclass[aps,prx,reprint,superscriptaddress]{revtex4-2}

\usepackage[document]{ragged2e}
\usepackage{xcolor}   
\usepackage{stackengine,graphicx}
\usepackage{amsmath}
\usepackage{etaremune}
\usepackage{braket}
\usepackage{mathtools}
\usepackage{anyfontsize}
\usepackage{appendix}
\usepackage{multirow}
\usepackage[export]{adjustbox}
\usepackage{placeins}
\usepackage{threeparttable} 
\usepackage{float}
\restylefloat{table}

\newcommand{\rev}[1]{{#1}} 

\definecolor{NewBlue}{rgb}{0, 0, 0.41}
\definecolor{NewRed}{rgb}{0.6, 0.07, 0.07}

\usepackage[colorlinks,
    linkcolor=NewBlue,
    citecolor=NewBlue,
    urlcolor=NewRed]{hyperref}

\newcommand{\EqualContrib}{These authors contributed equally to this work;  \href{mailto:choihr@mit.edu}{choihr@mit.edu}}

\begin{document}
\title{\textit{Tutorial:} Remote entanglement protocols for stationary qubits with photonic interface\rev{s}} 
\author{Hans K.C. Beukers}
\thanks{\EqualContrib}
\affiliation{QuTech, Delft University of Technology, PO Box 5046, 2600 GA Delft, The Netherlands}

\author{Matteo Pasini}
\thanks{\EqualContrib}
\affiliation{QuTech, Delft University of Technology, PO Box 5046, 2600 GA Delft, The Netherlands}

\author{Hyeongrak Choi}
\thanks{\EqualContrib}
\affiliation{Research Laboratory of Electronics, Massachusetts Institute of Technology, Cambridge, Massachusetts 02139, USA}

\author{Dirk Englund}
\email{englund@mit.edu}
\affiliation{Research Laboratory of Electronics, Massachusetts Institute of Technology, Cambridge, Massachusetts 02139, USA}

\author{Ronald Hanson}
\email{R.Hanson@tudelft.nl}
\affiliation{QuTech, Delft University of Technology, PO Box 5046, 2600 GA Delft, The Netherlands}

\author{Johannes Borregaard}
\email{J.Borregaard@tudelft.nl}
\affiliation{QuTech, Delft University of Technology, PO Box 5046, 2600 GA Delft, The Netherlands}

 
\begin{abstract} 
\justifying
\noindent Generating entanglement between distant quantum systems is at the core of quantum networking. \rev{In recent years, numerous theoretical protocols for remote entanglement generation have been proposed, of which many have been experimentally realized.} Here, we provide a modular theoretical framework to elucidate the general mechanisms of photon-mediated entanglement generation between single spins in atomic or solid-state systems. Our framework categorizes existing protocols at various levels of abstraction and allows for combining the elements of different schemes in new ways. These abstraction layers make it possible to readily compare protocols for different quantum hardware. To enable the practical evaluation of protocols tailored to specific experimental parameters, we have devised numerical simulations based on the framework with our codes available online.
\end{abstract}

\maketitle

\tableofcontents

\justifying
\newpage
\section*{Introduction}
Remote entanglement of quantum systems is a vital component in quantum networks and computing~\cite{kimble2008quantum, wehner2018_quantum_internet, ruf2021quantum}. Since remote stationary qubits cannot interact directly, a \rev{``}flying qubit" is needed to mediate the interaction and generate entanglement. Photons offer versatility, as they can transfer quantum information over long distances with low-loss optical fibers, operate at room temperature, and be easily detected with single-photon detectors. In a photon-mediated entanglement protocol, stationary qubits selectively interact with photons, usually via an optical transition, resulting in the entanglement between the photon and the stationary qubit. This entanglement between photons and spins can be used to create entanglement between two distant qubits. The entanglement generated can have high fidelity even in the presence of photonic loss, as photon detection can be used to herald a successful entanglement attempt~\cite{cabrillo1999creation, duan2001longdistance, browne2003robust, duan2003efficient, simon2003robust, barrett2005efficient, sangouard2007longdistance}\rev{, as long as sources of false heralding are limited}. These individual entanglement links between quantum nodes can then be combined to distribute the entanglement through a quantum network~\cite{pompili2021realization}. This distributed entanglement has been used to perform unconditional quantum teleportation between nodes without a direct optical link~\cite{hermans2022qubit}. \rev{Overall, these technologies pave the way toward network-based quantum computing~\cite{nickerson2014freely}, entanglement-based quantum communications~\cite{munro2015quantum}, distributed quantum sensing and long-distance interferometry~\cite{khabiboulline2019optical}.}

Photon-mediated remote entanglement of stationary qubits has been demonstrated in numerous quantum systems, including trapped ions~\cite{moehring2007_entanglement_ions}, neutral atoms~\cite{ritter2012_entanglement_atoms,hofmann2012_entanglement_atoms}, semiconductor quantum dots~\cite{delteil2016_entanglement_dots,stockill2017_entanglement_dots}, and color centers in diamond~\cite{bernien2013_entanglement,humphreys2018_deterministic}. Different demonstrations rely on different entanglement generation protocols, where the choice of protocol implementation is dictated by the features or limitations of the experimental platform. As a result, the implementation of the entanglement protocol is often tailored to specific hardware.

Here, \rev{as a \textit{Tutorial}}, we present a theoretical framework for comparing and understanding different photon-mediated remote entanglement protocols (REPs). The modular framework consists of four layers, with modules assembled by connecting one's output to another's input. 

The advantage of this framework is that it gives insights into the common features of remote entanglement protocols. Moreover, it allows easy modification of the modules to compare a protocol with different types of quantum hardware or to rearrange the quantum hardware and test different protocols with the same hardware (which is done in Section~\ref{sec:simulating}). Dedicated simulations of entanglement generation for a given experiment usually lack this flexibility.

We start by explaining the high-level idea and how the first layer (logical building block and topology) can describe REPs in a generic manner. Next, we introduce the encoding and physical building block layers, which link the protocol to specific hardware. Finally, the quantum optical modeling layer provides a detailed, quantitative description of the physical building blocks. Our focus is on qubits realized with a single spin in atomic or solid-state systems, but the framework can incorporate other systems such as the spin wave in an atomic ensemble and optomechanical resonators, as well as superconducting qubits. These single entangled links between qubits could later be extended to multi-spin encoding for error correction for fault-tolerant quantum networks~\cite{jiang2009quantum}.

\rev{We have released a software package titled QuREBB (Quantum Remote Entanglement Building Block simulations), available in an accompanying GitHub repository (Sec.~VI)~\cite{beukers2023qurebb}. This package offers a comparison of three exemplary entanglement protocols. For each protocol, we review every layer, from the idealized version to its practical implementation. As an example, we assume that the protocols are implemented with silicon-vacancy color centers in diamond nanophotonic cavities. This serves as a Tutorial on how to effectively employ our theoretical framework for the construction and quantitative analysis of the protocol's performance.} The modular structure of our software implementation mirrors the framework, enabling the implementation of complex quantum systems from the physical to the network levels. Users can easily add new physical devices or logical units without interfering with existing functionality. Our ``named quantum object'' simplifies the tensor operations of different quantum subsystems by indexing them by names instead of \rev{numerical values}.

Note that in each layer, we have descriptions of photons, spins, and spin-photon interfaces. In this work, we elaborate spin-photon interfaces in the most detail for their critical role in the creation of entanglement. We also systematically analyze the photonic and spin operations focusing on the construction of REPs. We refer readers interested in more details on spin or photon operations to the following references~\cite{dobrovitski2013quantum,flamini2018photonic, nielsen2010quantum}.

\begin{figure*}
    \centering
    \includegraphics[width=\textwidth]{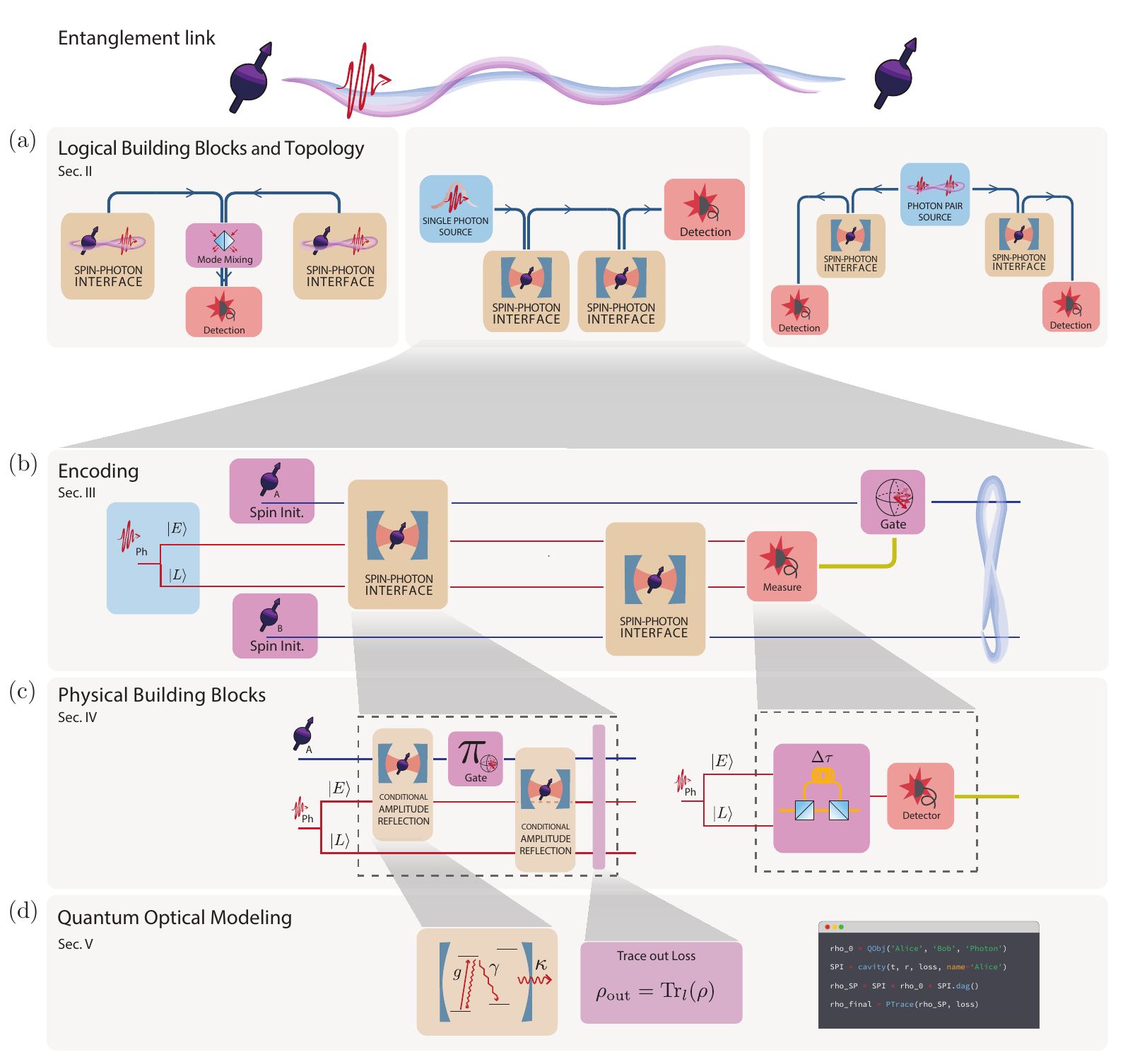}
    \caption{Overview of remote entanglement protocol modular framework. 
    \textbf{Entanglement link} - The objective of entanglement protocols is to entangle two remote spins (purple) where the entanglement is created by photons (red). 
    (a) \textbf{Logical Building Blocks and Topology} (Sec.~II) - The topology for remote entanglement generation is deconstructed in logical building blocks (LBBs). From left to right, we consider three main REP topologies: the detection-in-midpoint, sender-receiver, and source-in-midpoint topology.
    (b) \textbf{Encoding} (Sec.~III)- The encoding layer chooses the basis for the photonic qubits and the levels for the spin qubit. We illustrate the time-bin encoding of sender-receiver topology with spin-photon gates.
    (c) \textbf{Physical Building Blocks} (Sec.~IV) - The physical building blocks (PBBs) are the quantum channel describing the operations of the physical systems including imperfections. This layer shows the REP at the physical level. The first panel describes the construction of spin-photon gates with an overcoupled cavity QED system for state conditional amplitude reflection. The second panel shows the implementation of a measurement of the time-bin qubit in the X-basis, using a Mach-Zehnder interferometer (see Figs~\ref{fig:encoding} and \ref{fig:PBB photon}).
    (d)\textbf{Quantum optical modeling} (Sec.~V) - This layer models the hardware used in the PBB. In this example the relevant parameters for the PBBs (e.g. state-dependent reflection coefficient, $r_{\ket{0}, \ket{1}}$) can be calculated from physical variables (e.g. spin-cavity coupling rate, $g$), using a detailed quantum optical model. We implemented these descriptions in a software package publicly available on GitHub (Sec.~VI). Sec.~VII shows the simulation and benchmarking results with our framework.}
    \label{fig:overview}
\end{figure*}

\section{Remote entanglement protocol framework}

Figure~\ref{fig:overview} shows our modular framework for REPs. The framework consists of four layers, each becoming more specific and detailed in terms of hardware. Throughout the paper, we refer to stationary qubits as spins for convenience.

In the first layer, we choose the topology of the protocol and construct it using logical building blocks (LBBs) (Fig.~\ref{fig:overview}(a)). The topology, how the photons travel between the nodes, determines the generic high-level quantum circuit and the LBBs are the idealized quantum operations such as the spin-photon interface, the photon source, and the photodetector. The LBBs act on either the spin qubit, photonic qubit, or both. This layer simplifies and categorizes REPs, giving insight into the entanglement generation. The REP topology classifies into detection-in-midpoint, sender-receiver, and source-in-midpoint protocols~\cite{jones2016design}, as shown in Fig.~\ref{fig:overview}(a). We use the sender-receiver topology as an example in the figure.

In the second layer, we choose the qubit encoding, most importantly for the REP the photonic encoding: Fock state, polarization, dual rail, frequency encoding, or the time-bin encoding shown in Fig.~\ref{fig:overview}(b). This layer translates the abstract photonic and spin qubits into a specific implementation in the system.

In the third layer, we get closer to the hardware and implement the protocol in physical building blocks (PBBs). We implement the LBB with the desired hardware and encoding in physical building blocks as shown in Fig.~\ref{fig:overview}(c).  This requires the idealized quantum operations of the LBB to be compiled to the available PBBs. In Figure~\ref{fig:overview}, the spin-photon interface block is, for example, implemented as three PBBs, namely the reflection of the early time bin, a single-qubit rotation, and a reflection of the late time bin. The PBBs are a native operation for the hardware and are modeled as a quantum channel. 

The final layer is quantum optical modeling, where the exact physics of PBBs are modeled. This layer is added to reuse the quantum modeling of a system for different PBBs. For example, a critically or overcoupled cavity both require the same quantum optical modeling but are used in different PBBs.

\begin{figure*}
    \centering
    \includegraphics[width=0.95\textwidth]{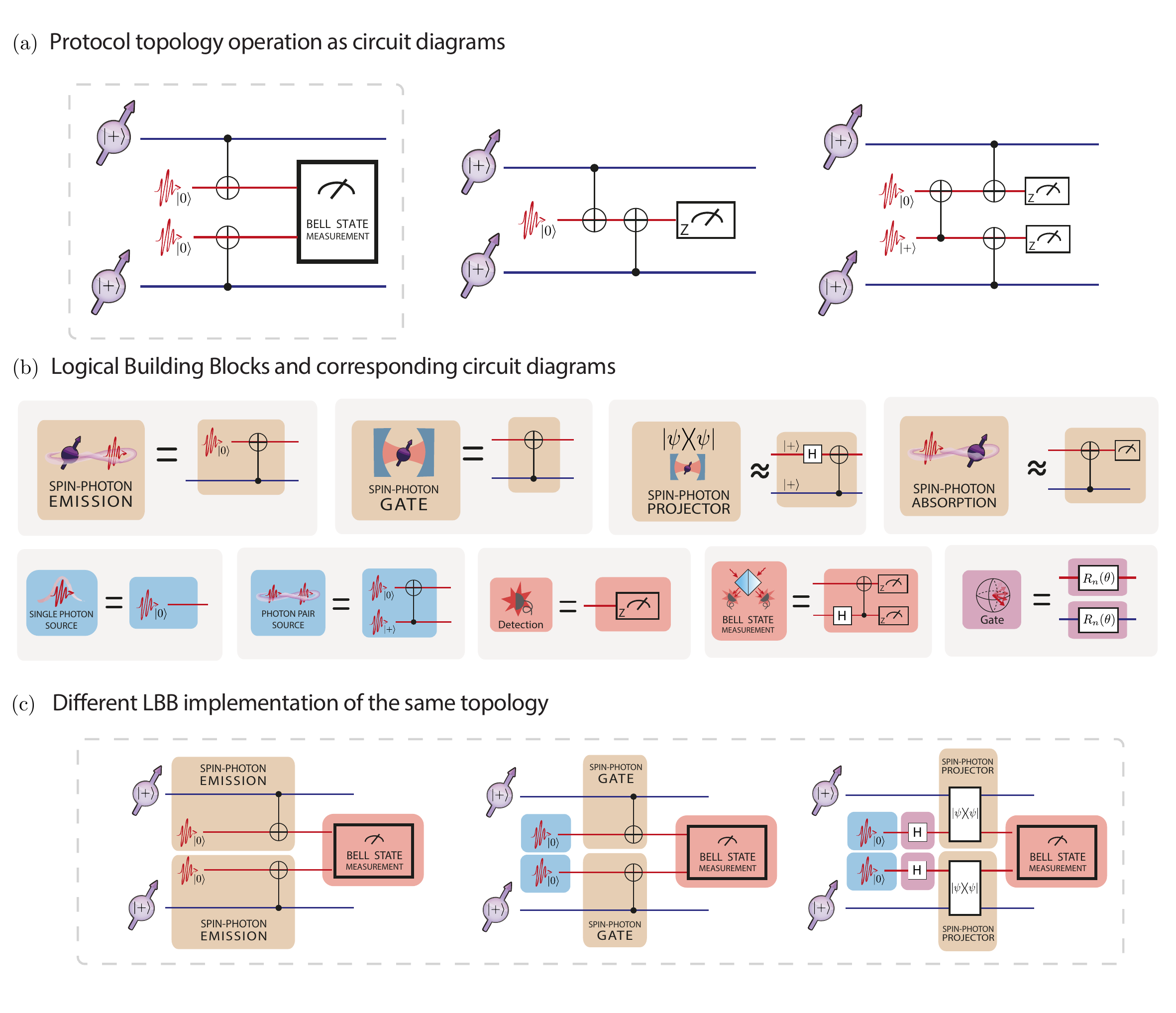}
    \caption{Remote entanglement protocol topologies and logical building blocks (LBBs): Figure (a) shows the three available topologies for REP - detection-in-midpoint, sender-receiver topology, and source-in-midpoint - and depicts their operation with circuit diagrams. \rev{Here we use the notation $\ket{+} = (\ket{0}+\ket{1})/\sqrt{2}$}. Figure (b) displays the LBBs available to construct the REP. The circuit description outlines their idealized operations, which the user can match with the diagrams in Figure (a). For the spin-photon projector and the spin-photon absorption, an approximate gate diagram is given. This can be used to see how they can be used in the quantum circuits in (a). The exact gate description for these LBBs would be a projector operation and a SWAP gate. Figure (c) provides three different implementations of the midpoint detector as an example. In this case, the spin-photon emission, spin-photon gate, and spin-photon projector are used as the spin-photon interface, respectively.}
\label{fig:LBB}
\end{figure*}

\section{Logical building blocks and Topology}
The first layer of our framework is the choice of protocol topology and its description in terms of logical building blocks (LBBs). \rev{The operation of the protocol topology can be described by the circuit diagrams (Fig.~\ref{fig:LBB}(a)).} LBBs are high-level quantum operations in a quantum network that can be chained together to form an entanglement protocol, such as photon sources, spin-photon emission, and photon detection. \rev{In Fig.~\ref{fig:LBB}(b) we show how LBBs compare to elementary circuit diagrams, and in Fig.~\ref{fig:LBB}(c) how they can be used to construct different protocol topologies.} The LBBs are not hardware-specific, as the practical details on what generates this operation and the dependence on physical parameters are described in the physical building block layer. 

\subsection{Topology of the protocol}
In this work, we consider three topologies for REPs, described in Fig.~\ref{fig:overview}(a), namely  detection-in-midpoint (ingoing),  sender-receiver, and source-in-midpoint (outgoing)~\cite{jones2016design}.

In the detection-in-midpoint topology, both endpoints generate spin-photon entanglement. The photons are then measured in the middle, in an entangled basis using a Bell state measurement. This projects the two spins in an entangled state. \rev{The detection-in-midpoint topology has an advantage in its simplicity. At its most basic, this topology requires only photon-emission from the spins, beamsplitter interference of the photons, and single-photon detections at the central station. Efficient spin-photon interfaces like optical cavities, while helpful, are not necessary to get high-fidelity entanglement. Furthermore, the time overhead for classical communication is reduced by half compared to the other topologies, since the endpoints communicate solely with the detection point, eliminating the need for direct communication between them.}

In the sender-receiver topology, the first endpoint generates spin-photon entanglement, and the photon is then sent to the second endpoint, where it interacts with the spin, such that entanglement between the spins is achieved. This interaction can be a gate-like behavior (e.g. by using an optical cavity) after which the photon is measured. Alternatively, the photon can be absorbed by the spin-photon interface in which case this topology matches the workings of quantum state transfer~\cite{ritter2012_entanglement_atoms, cirac1997quantum}. \rev{The sender-receiver topology does not necessitate an intermediary station for the spins to become entangled. However, the classical communication time for the heralding signal is twice as long compared to a detection-in-midpoint, as it needs to travel the full distance as opposed to half of it.}

The source-in-midpoint topology has an entangled photon source in the middle which sends these photons to the endpoints. The endpoints have an entangling interaction between the spin and the photon, after which the photons are measured and the spins will be projected into an entangled state. The source-in-midpoint topology is useful in satellite-assisted entanglement protocols. Ground-to-satellite channel (uplink) has a higher overall loss than the satellite-to-ground channel (downlink). Phenomenologically, this phenomenon is collectively referred to as the ``shower curtain effect'' of the atmosphere~\cite{dror1998experimental}. In particular, the diffraction and deflection by air turbulence are more severe in the early stage of transmission than in the late stage on the dispersed beam. Moreover, practical factors such as limited onboard optics on the satellite result in a larger pointing error in the uplink than the downlink~\cite{pirandola2021satellite}.

Remote entanglement between two stationary qubits has been experimentally realized with sender-receiver topology using trapped atoms embedded in optical cavities~\cite{daiss2021_distant_gate}, and with detection-in-midpoint using several platforms including NV centers in diamond~\cite{bernien2013_entanglement}, quantum dots~\cite{delteil2016_entanglement_dots}, trapped ions~\cite{moehring2007_entanglement_ions} and atoms~\cite{hofmann2012_entanglement_atoms}.\\

\subsection{Spin-Photon Interface blocks}
The key logical building block of a REP is the spin-photon interface block since this is the interaction between the stationary and flying qubits. 
In this work, we focus on REPs that are heralded, using single-photon detection to screen out photon loss errors. \rev{Absorption-based spin-photon interfaces generally allow heralding only through additional energy levels in the stationary qubit~\cite{ritter2012_entanglement_atoms}}. Thus, we give only a cursory treatment of absorption-based spin-photon interfaces. We identify four main categories of the spin-photon interface blocks:

\paragraph{Spin-Photon Emission} - A spin-photon emission block creates a photonic qubit entangled with the qubit state of the spin by emission through a higher level excited state. This is commonly achieved by the emission of a photon after spin-dependent excitation with a laser pulse. As an example, optical excitation of a spin in a superposition state will emit a photon depending on the spin state: $\frac{1}{\sqrt{2}}(\ket{0}_s+\ket{1}_s) \xrightarrow{\text{laser}} \frac{1}{\sqrt{2}}(\ket{0}_s+\ket{\text{excited}}_s) \rightarrow \frac{1}{\sqrt{2}}(\ket{0}_s\ket{0}_p+\ket{1}_s\ket{1}_p)$, \rev{where the subscript $s$ ($p$) labels the state of the spin (photonic mode), and $\ket{1}_s$ is the spin's bright state that is excited and emit a single photon.} Alternatively, the system can be brought to an excited state where two different decay channels lead to two different spin states. The emitted photons can have for example a different polarization entangled with the spin state.  Various optically active quantum systems have realized spin-photon emission through spin-dependent optical excitation or decay to different spin states, including NV centers in bulk diamonds~\cite{bernien2012_TPQI}, neutral atoms~\cite{chou2005measurementinduced}, trapped ions~\cite{moehring2007_entanglement_ions}, and quantum dots~\cite{delteil2016_entanglement_dots}.

\paragraph{Spin-Photon Gate} - A spin-photon gate block is a conditional gate between the spin and a photon. Depending on the exact implementation, this will act as a controlled Z- or controlled X-rotation on the photonic qubit. Spin-photon gate blocks require a strong coupling of photons with the stationary qubit, and this is realized by confining the light with photonic cavities or waveguides. In some cases, the need for strong coupling can be relaxed at the expense of a non-deterministic but heralded gate operation. Examples of practical implementation of spin-photon gate LBB are trapped atoms in Fabry-Perot cavities~\cite{daiss2021_distant_gate}, SiV centers in diamond photonic crystal cavities~\cite{bhaskar2020_repeater}, and quantum dots in photonic crystal waveguides~\cite{chan2022quantum}.
 
\paragraph{Spin-Photon Projector}
A spin-photon projector block selects only specific states from the input state. With the right input state and the right selection, this results in spin-photon entanglement. This method is sometimes referred to as "carving"~\cite{welte2017cavity}. An example is where the input state is in a superposition for both the photon and spin qubit (note that the $\ket{0}_p$ is referring to the qubit state, not the vacuum state): $\frac{1}{2}(\ket{0}_s\ket{0}_p+ \ket{0}_s\ket{1}_p+\ket{1}_s\ket{0}_p+ \ket{1}_s\ket{1}_p)$ and the spin-photon interface has a photonic loss that depends on both the spin and photon state. In this example, a spin-photon interface that has photon loss for the photon in the scenario's $\ket{0}_s\ket{1}_p $and $\ket{1}_s\ket{0}_p$ would create with 50\% chance the state $\frac{1}{2}(\ket{0}_s\ket{0}_p+ \ket{1}_s\ket{1}_p)$. So heralding no photon loss by photon detection ensures that only (an entangled) part of the incoming state is selected. This approach was used for entangling two neutral atoms in a cavity~\cite{welte2017cavity} and with silicon-vacancy centers in diamond photonic crystal cavities for the demonstration of an asynchronous Bell state measurement between two photons~\cite{bhaskar2020_repeater}. The downside is that it has intrinsic losses as it rejects part of the incoming state rather than performing a deterministic gate.

\paragraph{Spin-Photon Absorption}
The spin-photon absorption block transfers the photon state to the spin state. The block often implements a strong interaction using a cavity~\cite{reiserer2015cavitybased} or spin ensemble~\cite{heller2022raman}. The absorption of a photon from lossy channels results in the vacuum field and does not herald the entanglement by photodetection resulting in low fidelities. Instead, spin-photon absorption can be useful if carefully used. For example, one can read the spin state after the absorption, effectively constructing a heralded protocol. Alternatively, one can make a high optical-depth spin ensemble absorb a single photon from a photon-pair source and use the other photon from the source~\cite{piro2011heralded}. If the channel loss is negligible, the spin-photon absorption can directly implement quantum state transfer without classical communications~\cite{boozer2007reversible}.

\subsection{Photon blocks}

\paragraph{Photon source}
Photon source blocks can be considered as the initialization of photonic qubits. High-rate entanglement generation with photon sources requires the deterministic generation of single photons with high-efficiency quantum emitters~\cite{aharonovich2016solidstate,somaschi2016nearoptimal} or photon-pair generation followed by the heralding~\cite{mosley2008heralded}. For high fidelities, single photons need to be indistinguishable, which is often challenging in a solid-state environment~\cite{anderson2019electrical, grange2015cavityfunneled, choi2019cascaded}. The distinguishability problem is considered in the quantum optical modeling layer.

\paragraph{Photon pair sources}
Photon pair sources provide entangled photon pairs, which can be used for REPs with source-in-midpoint topology. This can be understood as two photons initialized in an entangled state. Mid-point on-demand entangled photon pair sources have been proposed based on correlated photon decay from quantum emitters~\cite{jones2016design}, mode-mixing of single photons~\cite{zhang2008demonstration} or multiplexing spontaneous pair sources~\cite{chen2023zeroaddedloss}.

\rev{We note that at the LBB level, photon pair sources output perfect Bell states in the logical basis, $\ket{0}$ and $\ket{1}$. The photonic basis of the states is determined in the photonic encoding layer (Sec.~\ref{sec:encoding}), and imperfections, such as probabilistic photon pair generation, are detailed in the quantum optical modeling of physical building blocks (Sec.~\ref{sec:QO}).}

\paragraph{Photon measurement}
Generally in the REPs, the photon is still entangled with the spins before detection. For example in the sender-receiver topology (middle circuit Fig~\ref{fig:LBB}a) the state before the measurement is 
\begin{equation*}
    \frac{1}{2}\left(\ket{0}_p(\ket{00}_{AB}+\ket{11}_{AB}) + \ket{1}_p(\ket{01}_{AB}+\ket{10}_{AB})\right),
\end{equation*}
\rev{where $A$ and $B$ refer to the two separate spins.}
This means that the photon needs to be measured on a basis that preserves entanglement between the spins, this is in the $Z$-basis for this situation. In this measurement of the photonic qubit, the outcome heralds a different entangled state on the spins, in this example measurement of $\ket{0}$ for the photon heralds $\frac{1}{\sqrt{2}}(\ket{00}+\ket{11})$. One can choose to feedback on one of the spins to always generate the same entangled state or use the measurement outcome in post-processing. 

\paragraph{Bell State Measurement}
\rev{Photonic Bell state measurement projects the state of two photons into one of four Bell states, $\ket{\Psi^\pm} = \ket{01}\pm\ket{10}, \ket{\Phi^\pm} = \ket{00}\pm\ket{11}$.} When it is required to have a quantum operation between two photons, for example in the detection-in-midpoint topology, the Bell state measurement can be used. This is advantageous as in linear optics the photons do not interact. This can be achieved with a beamsplitter and photon measurement after the beamsplitter. The Bell state measurement projects two photons into an entangled state. It is usually used for entanglement swapping of two spin-photon-entangled pairs to yield entangled spins. \rev{Note that the Bell state measurement with linear optics is probabilistic with only 50\% chance of succeeding. However, with auxiliary single photons, one can boost the success probability to 75\%~\cite{ewert2014efficient}}

\paragraph{Photon gates}
Gates on the photonic qubit are essential in some entanglement protocols. How easily these can be implemented completely depends on the photonic encoding used and this is discussed in Sec.\ref{ssec: PBB photon}.

\subsection{Spin blocks}
Operations on the spin such as initialization, gates, and measurements are their own logic building blocks. They are the standard set of initialization, qubit gates, and qubit measurement~\cite{nielsen2010quantum}. Entanglement protocols require in some cases the spin to be initialized and put in a superposition, for each entanglement attempt.

\subsection{Other logical building blocks}

The blocks discussed earlier are the logic operations required for the working of the REPs. On top of these there are processes, which in an ideal scenario do not do anything to the photonic and spin qubits. Their implementation will, however, add noise, for example, a photonic loss in a fiber or quantum frequency conversion of the photons. These blocks do not change the logical states of qubits in the ideal situation, so they appear as identity operators in the circuit diagrams of the LBB layer and they can be modeled in detail as PBBs. 

For example, a photonic loss block can be added, which takes care of the losses in the system, this does not change the ideal operation of the protocol but will impact the rate and fidelity in the simulations (which is implemented in the PBB). 

For a long-distance entanglement generation, it is advantageous to use photons in the telecom band to improve photon transmission using ultralow-loss fibers. As the physical platforms used for spin-photon interfaces have limited access to those wavelengths, quantum frequency conversion~\cite{degreve2012quantumdot, tchebotareva2019_telecom_entanglement} is often used to match the photon wavelength of the spin-photon interface with the desired communication band. In this case, one can model and add a frequency conversion block to protocols.

\subsection{Combining topology and Logical Building Blocks}
In a given topology, a user can implement different combinations of LBBs. Figure~\ref{fig:LBB}(c) demonstrates the detection-in-middle topology implemented with the spin-photon emission, gate, and projector logical building blocks.

\begin{figure}
    \centering
    \includegraphics[width=\linewidth]{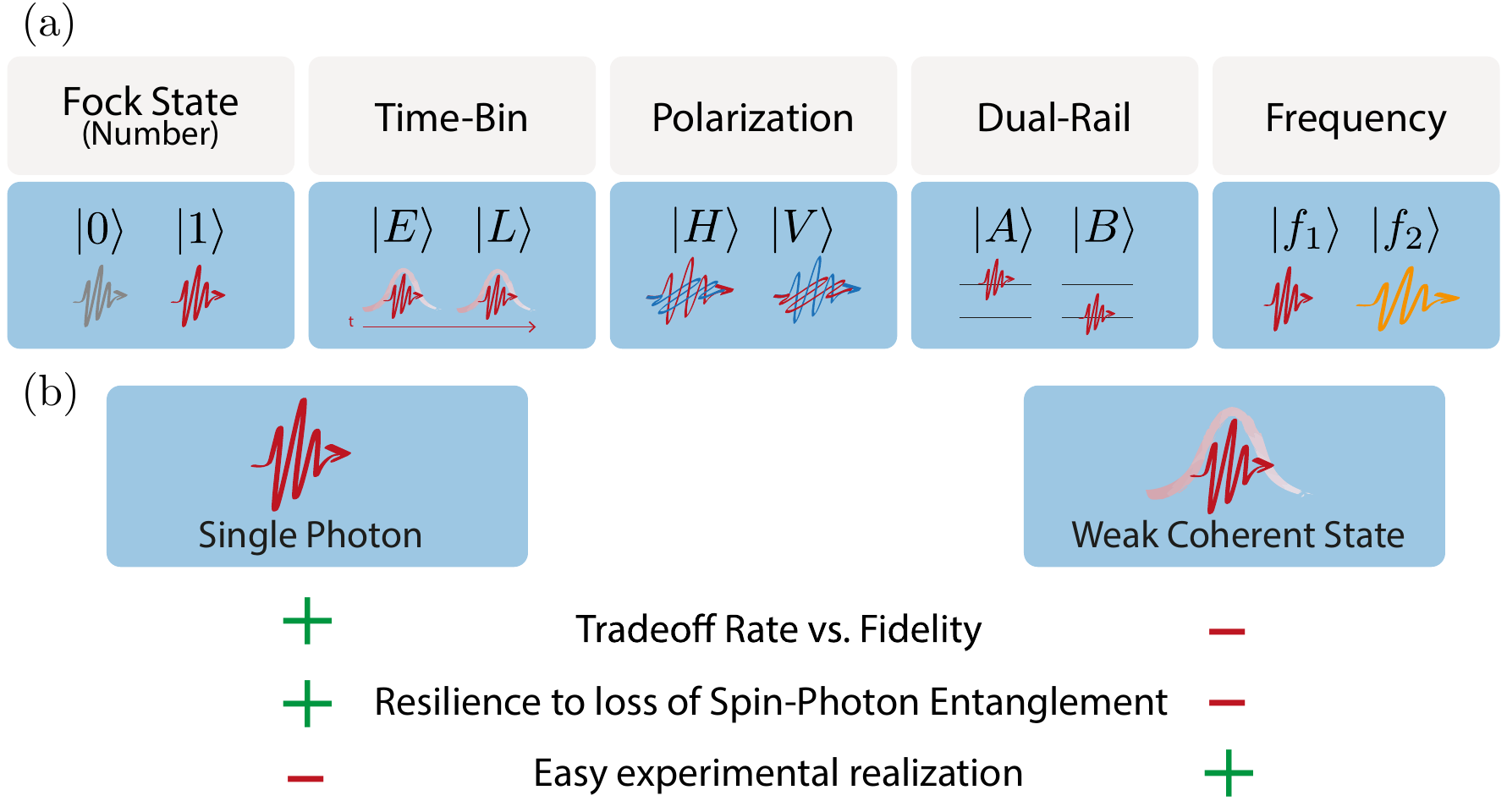}
    \caption{Encoding of photonic qubits. (a) Basis states of photonic encoding. (b) In every basis choice, the basis can be implemented with single photons or those can be approximated with a weak coherent state.}
    \label{fig:encoding}
\end{figure}

\section{Encoding}\label{sec:encoding}
In the previous section, the ideal operation of a REP was outlined. To translate this to the quantum hardware, the abstract spin and photonic qubits need to be encoded in the desired and available spin and photon levels. In this section, the photonic encodings are discussed in detail as these are very general. The spin encoding is much more platform dependent and therefore discussed more briefly.

\subsection{Photonic encoding}
\label{sec:photon_encoding}
Optical photons are the best option to send quantum information over a long distance for remote entanglement. Quantum information can be encoded into a photon using various degrees of freedom: amplitude (Fock-state encoding), timing (time-bin encoding), spatial modes (dual-rail encoding), polarization (polarization encoding), and frequency (frequency encoding); see Fig.~\ref{fig:encoding}. In literature, dual-rail encoding can also refer to general two-mode encodings e.g. time-bin and frequency encoding. In this tutorial, we will use the term dual-rail exclusively as spatial modes encoding. Each of the encoding schemes has two states acting as the qubit basis. In general, the other degrees of freedom are kept the same to simplify the operations and allow for the interference of encoded photons for a Bell measurement, which requires indistinguishable photons
unless precise measurement of conjugate variable is available and the measurement result is mutually unbiased (see e.g.,~\cite{metz2008effect, zhao2014entangling} for frequency mismatching case).

\subsubsection{Photon number of the photonic qubit}
All encoding except Fock-state encoding \rev{have} a single photon in the basis states. The advantage is that the loss of photons can be detected. The entanglement protocols discard the cases with photon loss for high-fidelity entanglement generation, so-called heralded entanglement~\cite{bernien2013_entanglement,humphreys2018_deterministic}.

\rev{As high-efficiency, indistinguishable single-photon sources are technologically demanding}, the single-photons are frequently replaced with weak coherent sources (attenuated lasers). They have Poisson statistics with a small mean photon number, well below one. The wave function of a weak coherent state is dominantly vacuum ($\ket{0}$) with a small fraction of single photon state ($\ket{1}$) and an even smaller fraction of two-photon states ($\ket{2}$):
\begin{equation*}
    \ket{\alpha} \propto \ket{0} + \alpha \ket{1} + \frac{\alpha^2}{\sqrt{2}} \ket{2} + ...
\end{equation*}
where $\alpha$ is the complex amplitude and $|\alpha|^2 << 1$ is the mean photon number. The vacuum component $\ket{0}$ reduces the rate of entanglement generation since it cannot herald entanglement through photodetection. The $\ket{2}$ state reduces the fidelity of the spin-photon entangled state as the loss of one of the two photons leaks information to the environment. There is, therefore, a trade-off between rate and fidelity for choosing $\alpha$, when using weak coherent states as approximate single-photon states in heralded entanglement generation protocols.

\subsubsection{Encoding basis}
\paragraph{Fock-state encoding}
\rev{Fock-state encoding stores quantum information in the photon-number eigenstates with zero and one photon ($\ket{n=0}$ and $\ket{n=1}$, where $n$ is the photon number). $n>1$ states are not considered in the encoding due to technical difficulties in preparing those states.} However, the problem is that loss takes one qubit state to the other and therefore directly impacts the fidelity (compared to other encodings where loss can be detected as the vacuum state is not part of the encoding space). The relative phase of the two bases evolves as the optical phase, so the optical path length needs to be stabilized or at least known for correction. \rev{Despite these complications, Fock-state encoding has a significant advantage in that Bell measurement is possible with the detection of a single photon, while the other encodings need the detection of two single photons.} The success probability of a heralded protocol scales linearly with the probability of photon loss if only single-photon detection is needed, while it scales quadratically for two-photon detection (in all but Fock state encoding). In cases where photon losses are significant, due to long-distance transmission or devices with low efficiency, the linear scaling given by Fock-state encoding can provide a key advantage over other encodings.

\paragraph{Time-bin encoding}
Time-bin encoding counteracts the drawbacks of Fock-state encoding at the expense of the requirement of two-photon detection for Bell measurement. Two time bins (early and late) are chosen to encode the photons. Photon loss can always be detected, in the ideal case of detectors with no dark counts, as this would result in no detection of photons. Therefore, the fidelity of protocols with time-bin encoding is not compromised, as long as dark counts are negligible with respect to the signal, but the rate decreases. Another advantage is relaxed phase stability: the optical phases should be stable on the time scale of the spacing of the time bins. However, arbitrary qubit operations are hard to implement, but the encoding is quite robust against noise sources such as dispersion or birefringence in the transmission medium. Therefore this encoding is mostly used in sending quantum information over long fibers and not in situations that require full control over the photonic qubit state~\cite{kok2007linear}.

\paragraph{Polarization}
The polarization encoding defines the qubit state in two perpendicular polarizations: horizontal and vertical polarization (H and V), diagonal and anti-diagonal (D and A), or left and right circular (L and R). Single-qubit gates are easily implemented, as all single-qubit rotations can be performed with waveplates (see Sec. \ref{ssec: PBB photon}). \rev{Moreover, polarization encoding requires phase stability between the two polarization bases. This stability can be readily attained in free space. In single-mode optical fibers the polarization is preserved but temperature and stress fluctuations in the fiber can rotate the polarization, stabilization and calibration is therefore required. Polarization maintaining (PM) fibers decouple the two different polarization bases by using orthogonal modes with different effective indices. This preserves only the amplitude in each basis and not the phase relation between them. In remote entanglement experiments using polarization encoding, usually single mode fibers are used~\cite{daiss2021_distant_gate, krutyanskiy2023entanglement}. }

\paragraph{Dual-rail encoding}
The dual-rail encoding uses two spatial modes for photons. The encoding has a main drawback as it requires twice the physical elements. The phase between two separate paths is extremely stable on the integrated photonic device, however, in fiber or free space the requirements are comparable to phase stabilization for Fock state encoding. This encoding can implement operations that are hard to perform in another encoding because path separation gives the most flexibility to use optical elements separately in two modes, see section~\ref{ssec: conversion}. 

\rev{Note that some literature uses the term 'dual-rail encoding' to refer to a single excitation out of two orthogonal bosonic degrees of freedom, encompassing spatial modes, polarizations, time bins, frequencies, wavevectors, and orbital angular momentum. In these instances, 'Fock state encoding' is referred to as 'single-rail encoding'~\cite{kok2007linear}. In this tutorial, we use 'dual-rail encoding' specifically for the single excitation in two spatial photonic modes and specify the degree of freedom in other cases.}

\paragraph{Frequency encoding}
In frequency encoding, photonic qubits utilize two distinct frequency modes as basis states. \rev{Single-qubit gates in frequency encoding require nonlinear optical devices such as electro-optic modulators that suffer from low efficiencies. To measure frequency-encoded photons, the frequency separation of two frequency modes must exceed the spectral resolution of a grating or a cavity. One can also directly detect a frequency-encoded photon in time-separated manner by using group velocity dispersion in optical fibers~\cite{pickston2021optimised} or Bragg gratings~\cite{davis2017pulsed}. Note that group velocity dispersion does not convert the frequency encoding to time-bin encoding. Even if a photon of one frequency is time-shifted relative to a photon of another frequency, the two photons still occupy distinct frequency bins.}

\subsection{Spin encoding}
Besides the photonic qubit, also the spin qubit needs to be encoded in the physical states of the system. The encoding of the stationary qubit depends on the system at hand. An important requirement for the spin-photon interface is that at least one of the states \rev{has an efficient and stable optical transition.} Besides, there should be ways to initialize, control, and read out with high fidelity. Lastly, the coherence time of the qubit should be long enough to bridge at least the time of flight of the photons to the midpoint or another quantum node as this allows for heralded entanglement generation, required for applications beyond point-to-point quantum key distribution~\cite{ruf2021quantum}. Examples of the levels that are used to encode the spin qubit in various systems are the electronic spin for NV centers in diamond~\cite{bernien2013_entanglement} or cold atoms~\cite{ritter2012_entanglement_atoms}, spin-orbital hybrid states for group-IV centers in diamond~\cite{bhaskar2020_repeater}, hyperfine levels in trapped ions~\cite{moehring2007_entanglement_ions} and angular momentum states in quantum dots~\cite{delteil2016_entanglement_dots}.

\section{Physical building blocks} 
\label{sec: PBB}
After the REP topology is constructed with LBBs and the encoding is chosen, we can construct the REP with the physical systems that are available. The LBBs are ideal circuit elements acting on the qubits of photons and spins, while the PBBs are physical processes on/of the optical modes (e.g. early and late) and spin states. Therefore, the PBB layer translates the abstract operations to allow hardware implementation. 

For example, to generate spin-photon entanglement, a spin-photon interface LBB can be composed of several PBBs that perform operations on different photonic states. As shown in Fig.~\ref{fig:overview}, a spin-photon interface LBB with time-bin encoding consists of a conditional \rev{amplitude} reflection PBB in early mode, qubit rotation PBB on the spin, and a conditional \rev{amplitude} reflection PBB acting again on the late mode. 

\begin{figure*}
    \centering
    \includegraphics[width=.9\textwidth]{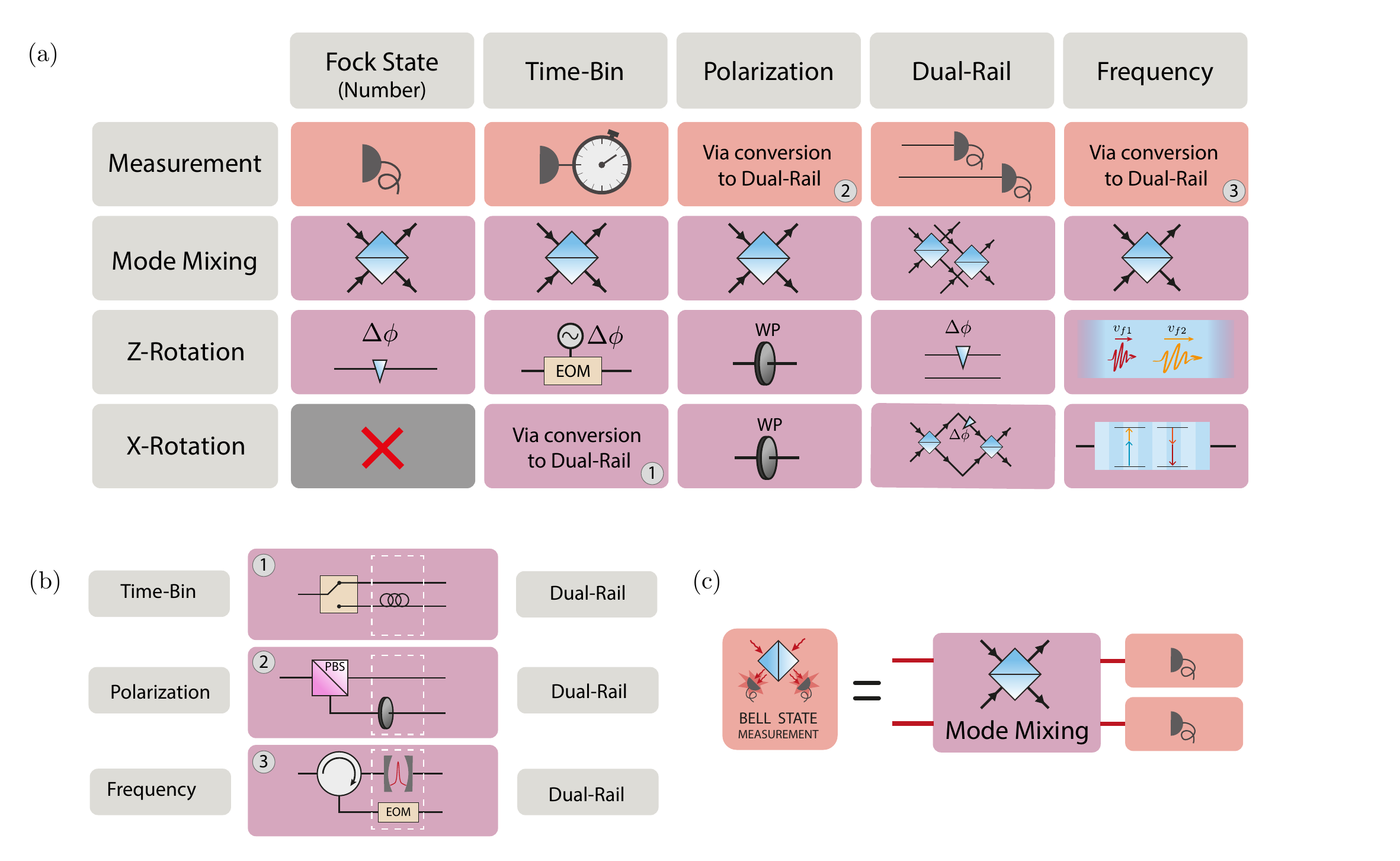}
    \caption{Photonic physical building blocks for the different photon encodings. (a) Each basis has its own implementation of the quantum operations, but some do not have natural implementation without converting them to a different basis (usually to dual-rail). The \textbf{measurement} in the computational basis is performed with a single photon detector. For time-bin encoding, it needs to be time-resolved. For polarization and frequency, the qubit is usually converted to and detected in dual-rail encoding. Two-qubit gates are implemented using \textbf{mode mixing} with a beamsplitter. The mode mixing lets two photonic qubits interfere with each other. For a \textbf{Z-rotation} the phase between the two basis states needs to be altered. This is implemented in Fock state encoding with a delay line, in time-bin encoding with an electric optic modulator (EOM) shifting one time-bin, in polarization encoding with a waveplate, in dual-rail encoding with delaying one of the lines, and in frequency encoding with a dispersive medium. The \textbf{X-rotation} requires changing the photon between the two eigenstates, which is not feasible for Fock states encoding, goes via dual-rail for time-bin encoding, can be easily implemented with a waveplate in polarization encoding, uses a Mach-Zehnder interferometer for dual-rail and nonlinear processes for frequency encoding. (b) The conversion between the basis that can be used for the easier physical implementation of quantum operations. Fock state encoding cannot easily be converted. All other encodings can be converted to and from dual-rail. For this, one element adds a new encoding (switch, PBS, or cavity/grating), and another removes the old encoding (in the white dashed box, delay, waveplate, or frequency shifting EOM). The latter can be omitted if the photon is detected afterward, as no indistinguishability is required. (c) The LBB of Bell State Measurement can be constructed using the mode mixing and single photon detection in each of the encodings.}
    \label{fig:PBB photon}
\end{figure*}

\subsection{Photon operations}
\label{ssec: PBB photon}
The PBBs of the photonic operations are described in Fig.~\ref{fig:PBB photon}. Polarization and dual-rail encoding have the convenience in the implementation of quantum gates and are often used in linear optics quantum computation~\cite{figgatt2019parallel}. The photonic PBBs, of which we describe photonic loss, mode mixing, and photodetection in Sec.~\ref{ssec: quantum channel}, are often studied in standard quantum optics textbooks.

\subsubsection{Measurement}
Measurement in all photon encodings is done by means of single photon detectors. Fock state encoding can be measured directly. \rev{However, a single photon state ($\ket{1}$) after photon loss cannot be distinguished from the vacuum state ($\ket{0}$).} Time-bin encoding requires time-resolving detectors. Polarization and frequency-encoded qubits cannot be detected directly and are usually converted to dual-rail (see Sec~\ref{ssec: conversion}) where both modes are then measured with a separate detector.

\subsubsection{Photon gates}
An arbitrary gate can be made with the combination of rotations around the X- and Z-axis of the qubit Bloch sphere. For a Z-rotation, the phase between the qubit basis states needs to be changed. This is directly also the phase that needs to be stable for the use of the encoding. For an X-rotation, the operation needs to change the basis states. This can be readily implemented for polarization encoding with waveplates. \rev{For a Fock state encoding, this is not trivial and has not been demonstrated to the best of our knowledge,} and the time-bin needs to be converted to dual-rail. In dual-rail encoding a Mach-Zehnder interferometer is used.

\subsubsection{Bell state measurement}
The Bell state measurement can be performed with linear optics with a 50\% success probability~\cite{kok2007linear}. It can only detect the $\ket{\Psi^\pm} = \ket{01} \pm \ket{10}$ as it requires detecting both modes after the beamsplitter. The $\ket{\Phi^\pm} = \ket{00} \pm \ket{11}$ cannot be detected as measuring the same modes after the beamsplitter reveals both their individual states. The Bell state measurement is done by using a beamsplitter for mode mixing and two detectors to tell which of the two detectable entangled states ($\ket{\Psi^+}$ or $\ket{\Psi^-}$) was measured (Fig~\ref{fig:PBB photon}(c)).

\subsubsection{Conversion between encoding}
\label{ssec: conversion}
Figure~\ref{fig:PBB photon}(b) shows the possible conversion between the encoding. In practice, dual-rail encoding is \rev{a versatile} encoding that can be converted directly into the others and vice versa. For the conversion, one needs the separation of the photons in different modes into different spatial paths (e.g. using a polarizing beamsplitter) and compensating elements to remove the character from the previous encoding (e.g. using a waveplate). The compensation can be omitted if one does not need indistinguishable photons after conversion, for example, if the photons are detected directly.

The conversion between time-bin encoding and dual rail is used often as this is the only viable way to perform X-gates on the photonic qubit. In this case, the switch is often replaced by a beamsplitter by accepting the 50\% loss, as this reduces the complexity of the experiment and as the time delay between time bins can be short compared to the switching time. When using a beamsplitter, there are also possible optical paths involving the unwanted output ports that cause the photon to be delayed and fall outside of the defined time-bin window, however, these cases can be discarded upon measurement using time-resolving detectors.

\subsection{Spin operations}
The exact PBB spin operations depend a lot on the chosen spin encoding. We discuss the most common ones below.

\subsubsection{Initialization}
Initialization of the spin is usually achieved in one of the following ways. When a non-perfectly cycling optical transition is present, which means that under continuous driving there is a (small) probability of spontaneously decaying to a different state than the ones involved in the transition, it is possible to use optical pumping~\cite{reiserer2016_subspace}. By laser excitation of the non-cycling transition, the system will eventually decay to the desired qubit state and remain there as this is not driven by the laser. In systems with a very good readout, it can be advantageous to read and generate the desired initialized state with a control pulse conditional on the readout result~\cite{bhaskar2020_repeater}.
If one wants to prepare the system in a superposition a quantum gate can be used. Alternatively, one can prepare a state by Stimulated Raman Adiabatic Passage (STIRAP)~\cite{bergmann2015perspective}. 

\subsubsection{Control}
Quantum control of the qubit is usually achieved by direct Rabi driving of the transition. This is usually a microwave~\cite{dobrovitski2013quantum} or optical field but can also be done with other coupled fields, like an oscillating strain field~\cite{maity2020coherent}. 
If the transition frequency is experimentally hard to reach or it is only weakly allowed, two driving fields in a lambda configuration can be used, in which the qubit states are coupled to a common excited state. By driving both transitions, a coherent rotation on the qubit can be achieved with a two-photon Raman transition~\cite{pingault2017_MW_control}.
In some systems, mostly based on solid-state implementations, coupling to fluctuating magnetic fields in the surrounding environment such as nuclear spin baths can affect the coherence of spin-based qubits. With the control in place, it is usually possible to extend the coherence time of the spin qubit in these platforms by means of dynamical decoupling~\cite{delange2010_decoupling}.

\subsubsection{Readout}
The qubits we are considering have a good spin-photon interface, which allows for optical readout of the spin. The spin can read out by the detection of the state-dependent fluorescence. For high fidelity of this readout, high collection efficiency and good cyclicity of the optical transition are needed. A good cyclicity results in a transition that can be driven for a long time generating as many photons as possible before the qubit decays to an unwanted state~\cite{nguyen2019_PRL}, a high collection efficiency allows for measuring enough photons for a high-fidelity readout with a shorter driving time and lower probability of the qubit changing state. When a conditional phase reflection PBB (see next section) is used the spin state can be read out by phase readout~\cite{stas2022robust}.

\subsubsection{Auxiliary qubits}
To go beyond a single entangled link it is essential to have auxiliary qubits for storage of quantum states in the system~\cite{childress2005faulttolerant}. This is for example required for entanglement distillation~\cite{bennett1996purification,deutsch1996quantum,campbell2008measurementbased,kalb2017_distillation} and entanglement swapping in repeater schemes. The available auxiliary qubits are highly dependent on the system and can go from nuclear spins in solid-state emitters to different species of atoms or ions in those systems. The most difficult requirement to fulfill for these qubits is often that they need to be resilient for entanglement attempts on the spin qubit, which typically involves a lot of initialization and control pulses. This can result in a competition between the coupling required for two-qubit gates and the isolation required for resilience during entanglement attempts~\cite{reiserer2016_subspace}.

\begin{figure*}
    \centering
    \includegraphics[width=0.72\textwidth,center]{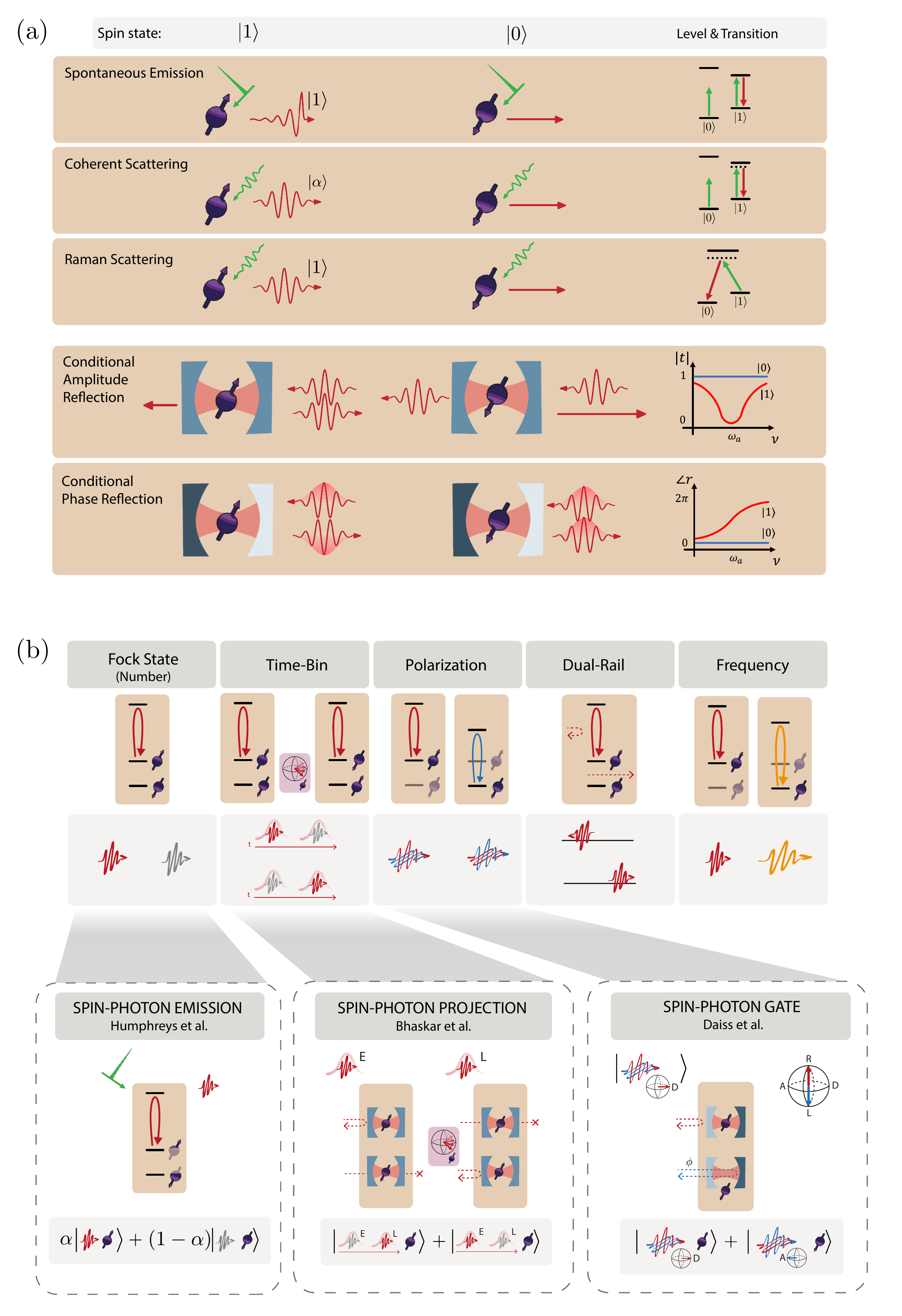}\\  
    \caption{ Spin-photon interface physical building blocks - (a) State-selective optical transition as spin-photon interfaces. For the \textbf{spontaneous emission} spin-photon interface the excitation pulse is much shorter than the optical lifetime and excites the $\ket{1}$, it will emit a single photon by spontaneous emission. The \textbf{coherent scattering} spin-photon interface scatters a weak pulse from the $\ket{1}$ optical transition, resulting in a photon with a weak coherent state. The \textbf{Raman scattering} spin-photon interface is excited with a pulse but as the optical excitation ends up in a different spin state only a single photon is emitted. In the \textbf{conditional phase reflection} spin-photon interface a photon is reflected from the cavity with a coupled spin. For $\ket{1}$ the phase is flipped and therefore entangled with the spin state. An overcoupled cavity is used for this purpose. In the \textbf{conditional amplitude reflection} spin-photon interface the photon is transmitted or reflected depending on the spin state. A critically coupled cavity is used for this case. (b) Schematic example of how a spin-photon interface LBB can be implemented in the different photonic encodings starting from a spin-photon PBB. Below are examples of experimental realizations of the emission~\cite{humphreys2018_deterministic}, projection~\cite{bhaskar2020_repeater}, and gate~\cite{daiss2021_distant_gate} LBBs in different encodings.}
    \label{fig:PBB}
\end{figure*}

\subsection{Spin-photon interface}
The spin-photon interface is an essential part of the entanglement protocols as it connects the stationary and flying qubits. We discuss five different spin-photon interface PBBs, see Fig~\ref{fig:PBB}(a). The first three are suitable to make spin photon entanglement for a spin-photon emission LBB: spontaneous emission, coherent scattering, and Raman scattering. These PBBs are all controlled with a classical laser control pulse and the output is a single photon (or a weak coherent state depending on the exact implementation) entangled with the spin. 
When the interaction between the light and the emitter is very strong, we can get a reflection of a single photon or weak coherent state where the amplitude or phase of the reflected light is controlled by the spin state. The conditional amplitude or phase reflection PBB can be used to build the LBB of a spin-photon projection or a spin-photon gate.

\subsubsection{Spontaneous emission}
The conceptually most simple way to implement a spin-photon interface is the emission-based spin-photon interface where the spin-photon entanglement is generated through spontaneous emission. In this case, a short, high-power, optical $\pi$ pulse is used to excite the system to the excited state. If this excitation or the spontaneous emission depends on the spin state we can create spin-photon \rev{entanglement}.

With this PBB we can create a spin-photon emission LBB. For Fock state encoding we prepare the spin in a superposition and apply a spin state-dependent optical $\pi$-pulse such that the presence of a photon is entangled with only one of the spin states. For the time-bin encoding, we initialize the spin state in an equal superposition and apply the excitation twice with a $\pi$-pulse of the spin in the middle. For the polarization of frequency encoding, both spin states need to have an optical transition with a different polarization or frequency. The entanglement can either be generated by initializing the spin in an equal superposition and exciting both transitions or by relying on an optical transition that has an equal chance of decaying to either state correlated with a different polarization or frequency. An example of polarization encoding can be found in~\cite{wilk2007singleatom}.

The photon coming from such a process will have frequency and exponential temporal shape determined by the optical properties of the emitter. The linewidth is determined by the lifetime and the inhomogeneously broadened linewidth and the indistinguishability is determined by the dephasing processes of the optical transition. This requires favorable properties of the emitter to make the photons suitable for entanglement generation.
The optical $\pi$-pulse needs to be much shorter than the optical lifetime, otherwise in case of fast spontaneous decay the same pulse can cause re-excitation of the transition and a second photon emission, which can lead to an error in the protocol. On the other hand, it needs to be not too short as this increases the linewidth of the excitation pulse and can therefore couple to and cause photon emission from unwanted off-resonant transitions which are close in frequency.
A common method to separate the single photon from the excitation laser is to use the delayed emission of the photons compared to the arrival of the laser pulse. By delaying the detection window compared to the arrival of the excitation pulse one can filter out the laser pulse in the time domain.

This PBB was used to make entanglement between NV centers~\cite{humphreys2018_deterministic}, see Fig~\ref{fig:PBB}(b).

\subsubsection{Coherent scattering}
In the coherent scattering PBB a long and weak pulse is scattered elastically from the emitter. Under the condition that the exciting pulse is weak and long enough that on average only one photon interacts with the emitter within the optical lifetime, the scattered light will inherit the temporal shape, photon statistics, and frequency of the excitation field~\cite{childress2005faulttolerant}.
One can detune the laser from the emitter, making it more resilient to the spectral instability of the emitter. However, as the light has the same character as the laser, filtering the excitation light is experimentally much more challenging. This PBB allows for a spin-photon emission LBB in a similar way as the spontaneous emission.

\subsubsection{Raman scattering}
The third way to make a spin-photon emission LBB is using the PBB of Raman scattering. In this case, the driving field and the emitted photon are coupled in a lambda scheme where they both couple to a virtual state (see Fig~\ref{fig:PBB} third row). In this inelastic scattering event, the temporal waveform and frequency are determined by the driving field, but now only a single photon is emitted. The driving field and the spontaneously emitted photons, which would act as noise sources, can be separated and filtered out from the scattered photons since they have different frequencies. There is usually a trade-off between efficiency and noise in this PBB by choosing the detuning of the virtual level from the excited state. This PBB was used for entangling trapped ions over hundreds of meters~\cite{krutyanskiy2023entanglement}. 

Raman transition can be also used for spin-photon absorption. In this case, the spin absorbs the incoming photon with one transition, and the other transition is driven so that the spin is in another state upon absorption. If there is no photon interacting with the first transition, for instance because there is no photon or it is in a different polarization, then there is no change in the spin state because the driving addresses the transition with no population.

\subsubsection{Conditional amplitude reflection}
When the interaction between the spin and the photon becomes strong, a single spin can modulate the single photon. This is different from the three previous situations where the driving field is 'classical' and only the photon emitted or scattered is in the single photon regime. As emitters have a limited dipole, the light field needs to be confined spatially to achieve strong interaction, which can be achieved in an optical cavity or a waveguide. 
In the conditional amplitude reflection PBB, the emitter acts as a quantum switch: reflecting the light if it is in one spin state while transmitting the light in the other spin state.

In the dual rail encoding, this can clearly be used to realize a spin-photon gate as the spin interacts directly with the two spatial paths. For the other photonic encodings, we can use this PBB to make the spin-photon projector LBB: the photon is only reflected if the transition is on resonance and otherwise lost. This PBB was used to show an asynchronous Bell state measurement with the silicon vacancy in diamond~\cite{bhaskar2020_repeater}, where a time-bin photonic qubit and a nanophotonic optical cavity were used, see Fig~\ref{fig:PBB}(b).

\subsubsection{Conditional phase reflection}
To perform a spin-photon gate the conditional amplitude reflection PBB works only for the dual-rail encoding. For the other encodings, the problem is that photons are lost in transmission depending on the spin state, which is not compatible with a spin-photon gate. To apply a spin-photon gate we need an unconditional reflection of the photon. The conditionality of the interaction should therefore be encoded in the phase of the photon. This is achieved with an overcoupled cavity where the back side of the cavity has a much higher reflectivity than the front such that the photon will not be transmitted. The emitter will alter the cavity response in such a way that depending on the spin state, the photon will be reflected on the cavity or will enter the cavity and leave it again. When calibrated well, this can result in a $\pi$ phase shift difference between the two types of reflection. The cavity can in principle also be replaced by a waveguide with a mirror at the end.

The conditional phase reflection with polarization-encoded photonic qubits was the central PBB in the realization of a non-local gate with cold atoms in cavities~\cite{daiss2021_distant_gate} (see Fig~\ref{fig:PBB}(b)). 

\begin{figure*}
    \centering
    \includegraphics[width=\textwidth]{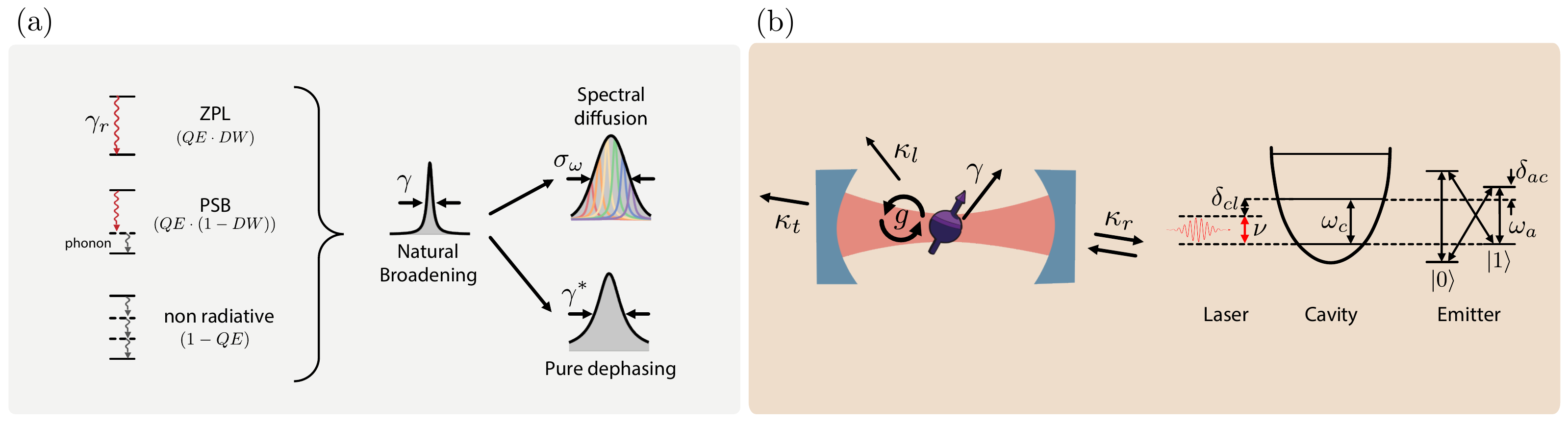}
    \caption{ \textbf{Quantum Modeling layer - spin-photon interface.} (a) The inefficiencies and noises of the optical transition in the spin-photon interface. (Left) zero-phonon-line decay ($\gamma_r$), phonon-sideband decay, and non-radiative decay combined give a total decay rate $\gamma$ resulting in the natural broadening (Lorentzian). The ratios of transitions are expressed with an internal quantum efficiency ($QE$) and Debye-Waller factor ($DW$). (Right) Additional slow noise adds Gaussian spectral diffusion with a standard deviation of $\sigma_\omega$, while the fast noise gives Lorentzian pure dephasing of linewidth $\gamma^*$. (b) (Left) cavity-QED system parameters. $g$ is vacuum Rabi frequency, $\gamma$ the spontaneous emission rate, and $\kappa_r, \kappa_t,$ and $\kappa_l$ are cavity decay rate to reflection, transmission, and loss port, respectively. (Right) level diagram of laser, cavity, and spin. We assume that $\ket{1}$ state of spins is desirably interacting with photons; interaction with the $\ket{0}$ state is negligible when $\gamma \ll \delta_{01}, g$.}
    \label{fig:QOModeling}
\end{figure*}

\section{Quantum optical modeling}
\label{sec:QO}
In the quantum optical modeling layer, the devices used are simulated at the quantum optics level. The aim of this layer is to calculate the action of the PBB on the photonic modes and spin states. The operations of PBBs are modeled as quantum channels which are completely positive trace-preserving maps. PBBs map an input density matrix to an output density matrix. We emphasize that the mathematical modeling of PBBs can be done in different levels of detail. Thus, one can choose the PBB model as exhaustive or approximate depending on the purpose. The quantum optical modeling layer is separated from the PBB layer as many PBBs rely on the same physical systems and therefore the same modeling, e.g. the conditional amplitude and phase reflection PBB both use an emitter coupled to a cavity. Moreover, the same PBBs can be modeled in different ways with different details. We will first discuss the intrinsic properties of the emitter, then how they can be enhanced and modified by an optical cavity, and finally, the modeling of a couple of PBBs.

Using the modeling of a hardware-specific implementation, a REP can be benchmarked and optimized with respect to relevant experimental parameters. Having a realistic description of the building blocks also allows us to take into account noise sources and practical limitations.

\subsection{Emitter}
For the spin photon interface to perform well, the properties of the optical interface are important. The optical transition of an ideal emitter (Fig.~\ref{fig:QOModeling}(a)) has a naturally broadened linewidth $\gamma$ related to its excited state lifetime. In practice, noises in the surrounding environment of the emitter generally cause the optical line to broaden~\cite{choi2019cascaded,wolfowicz2021quantum}. 

The noises can be divided into spectral diffusion and dephasing by their time scales. Especially for solid-state emitters, the fluctuation of the local charge environment is often slower than the nanoseconds' lifetime of the transition. These charge fluctuations can change the spacing of the energy levels involved in the optical transition by Stark effect, causing the frequency to change. This effect is called spectral diffusion. Spectrally diffused transitions have fluctuating resonance frequency, shot-to-shot, that can be modeled with a Gaussian distribution of standard deviation $\sigma_\omega$. The fidelity of spectrally diffused spin-photon interfaces can be calculated by a Monte Carlo simulation, statistically averaging over the distribution function~\cite{choi2019cascaded}.
On the other hand, fast noise sources such as mobile charges on the surfaces of photonic nanostructures or acoustic phonon scattering induce homogeneous broadening of the transition. In this limit, the frequency and phase noises are indistinguishable, $S_{\delta\omega_1}(\omega) = S_{\delta\phi_1}(\omega)\cdot \omega^2$ where $S(\omega)$ is the power spectral density, and the subscript of $S$ refers the random process. The noise is modeled as random phase flips, or equivalently, pure-dephasing without longitudinal relaxation. For pure dephasing $\gamma^*$, the transition is further broadened by $\gamma^*$ from radiative broadening, and the lineshape is Lorentzian. \rev{If the fluctuation time scale of a noise is in an intermediate regime, the dynamics of the spin-photon interfaces and the resulting fidelity of entanglement requires a complete specification of the noise as a stochastic random process.}

Furthermore, especially in solid-state emitters, in addition to the coherent transition without the involvement of phonons: the zero-phonon line, there are phonon-assisted optical transitions, called phonon sidebands. Moreover, non-radiative, multiphonon processes make the excited-state population decay without emitting photons. Quantum efficiency ($QE$) is defined as the ratio between the rates of all radiative processes to those of all processes:
\begin{align}
    QE &= \frac{\Gamma_\textrm{rad}}{\Gamma_\textrm{rad} + \Gamma_\textrm{non-rad}}.
\end{align}
The decay rate of the zero-phonon line relative to the phonon sideband is the Debye-Waller factor ($DW$). In remote entanglement settings, a large $QE\cdot DW$ is helpful for an efficient spin-photon interface, since only the coherent photons of the zero-phonon line are usable for entanglement protocols.

Other noise sources, such as a magnetic field or strain fluctuations, can also influence the transition through Zeeman shifts or mechanical deformation of the substrate, altering the energy levels. Moreover, in platforms other than solid-state emitters, like trapped ions or superconducting qubits, other sources of noise such as thermal fluctuations or coupling to nearby qubits can play a significant role.

\subsection{Photonic cavity}
\label{subsec:cavity}
A bare emitter can be used as a spin-photon interface, but it is advantageous to enhance the emitter properties and the collection efficiency of photons by means of an optical cavity. Other devices can be used to achieve the same goal, such as coupling the emitter to (photonic-crystal) waveguides or optical fibers. However, using a cavity is the most common approach and often other devices can be described as a special case of cavity-emitter coupling~\cite{kiilerich2019inputoutput,kiilerich2020quantum}.
Quantum optical modeling of a two-level system coupled to an optical cavity is a well-studied subject~\cite{reiserer2015cavitybased}. \rev{See also Ref.~\cite{saleh2019chapter, yariv2007photonics} for various cavity designs and the operations of cavities, including ring resonators~\cite{ngan2023quantum}.}

Figure~\ref{fig:QOModeling}(b) shows relevant quantities of the emitter-cavity system. The dynamics of an emitter-cavity coupled system are governed by three parameters, $\gamma$, $\kappa$, and $g$. The emitter (cavity) transition linewidth $\gamma$ ($\kappa$) is the inverse time of emitter decay \rev{(cavity relaxation)}. Spontaneously emitted photons into free space, non-radiative decays, and decays into phonon sidebands fall outside the coherent cavity-emitter interactions and are considered as a relaxation channel of the system, incorporated in $\gamma$. The dynamics of a cavity-emitter coupled system are governed by the coupling rate $g$. Physically, $g$ is the Rabi-frequency when the emitter is driven by a vacuum cavity electric field (zero-point electric field multiplied by the transition dipole moment).  

The cooperativity of the system is defined as:
\begin{align}
    C &= \frac{4 g^2}{\kappa \gamma}.
    \label{eq:cooperativity}
\end{align}
The cooperativity gauges the strength of coherent interaction relative to the dissipation, and higher cooperativity improves the efficiency of spin-photon interfaces. In REPs, this directly translates to the rate and the fidelities, as we will see in Sec. VII.

While $\kappa, g$, and $\gamma$ are usually set parameters once the device is realized, the operation point of the cavity-emitter system can often be tuned. In the right side of Fig.~\ref{fig:QOModeling}(b) we indicate the energy relations that determine the operation point. The frequencies $\omega_c, \omega_a$, and $\nu$ are related to the energies of the cavity mode, atomic transition, and external source (laser or single photon). The detuning between the cavity and emitter, $\delta_{ac}$, determines how the emitter and cavity couple and what is the amplitude and phase response of the system. The external source frequency, which we define with respect to the cavity as $\delta_{cl}$, can be optimized to obtain the desired phase or amplitude when interacting with the cavity.
Realistically, there are often additional optical transitions that couple (or are close) to the cavity. Here, we depict a common level scheme for solid-state emitters, the Voigt configuration. When the energy difference between the target transition (here involving the bright state $|1\rangle$) and other transitions is comparable to the cavity linewidth, emitter linewidth, or coupling rate, or the input light is not far-detuned, undesired interaction of the dark $\ket{0}$ state can lead to errors.

The cavity loss channels, $\kappa_{t, r, l}$, determine the cavity behavior. It is useful to relate the losses to the cavity input channel, which we take to be $\kappa_r$, in this way, we can define the cavity output coupling as critically coupled ($\kappa_l + \kappa_t = \kappa_r$), undercoupled ($\kappa_l+\kappa_t > \kappa_r$) or overcoupled ($\kappa_l + \kappa_t < \kappa_r$). The first and latter regimes are particularly interesting for a cavity-based spin-photon interface: an over-coupled cavity is sometimes called a single-sided cavity as the preferential output channel is the one used for probing the cavity; therefore emitted photons will be funneled in the collected optical mode, while incident photon will almost always be reflected back to the same port; this cavity is used in the conditional phase reflection PBB. A critically-coupled cavity allows both transmission and reflection to potentially be used as output channels, this is the cavity used for a conditional amplitude reflection PBB.

Emission-based spin-photon interfaces can be realized using a bare emitter, though this can pose a severe limitation to the entanglement generation rate when quantum efficiencies and/or Debye-Waller factors are low. In addition, especially for the spontaneous emission, spectral diffusion and dephasing directly affect the photon indistinguishability and thus the entanglement fidelity. The efficiency and the fidelity of a spin-photon interface can be improved with a cavity: the Purcell effect enhances the emission of the optical transition by decreasing the lifetime and increasing the fraction of photons emitted by the target transition (e.g. zero-phonon line for solid-state spins). At the same time, the optical linewidth broadening coming from the lifetime reduction decreases the relative effect of incoherent broadening mechanisms. The Purcell factor is defined as;
\begin{align}
F_p = 4g^2/\kappa\gamma_r = \frac{3}{4\pi^2}\frac{Q}{V}\left(\frac{\lambda}{n}\right)^3,
\end{align}
where $Q$ is the cavity quality factor, $\lambda$ is the resonant wavelength, $n$ is the refractive index of the host material, and $V$ the effective mode volume. The effective mode volume is calculated with the electric field profile of cavity mode, $E(\vec{r})$; 
\begin{align}
    V = \int dV \epsilon(\vec{r})|E(\vec{r})|^2/\epsilon(\vec{r}_e)|E(\vec{r}_e)|^2,
\end{align} 
for the emitter located at $\vec{r}_e$, and $\epsilon(\vec{r})$ is the permittivity as a function of coordinates~\cite{reiserer2015cavitybased,choi2017selfsimilar}. \rev{If the cavity has sufficiently high quality factor and small mode volume, resulting in $g>\kappa, \gamma$, the system enters the strong coupling regime~\cite{reiserer2015cavitybased}}, and the overall outcoupling efficiency of the emitted photons can become inefficient. This is because the cavity photon can be reabsorbed into the emitter before escaping the cavity. In the strong coupling regime, the outcoupling efficiency is,
\begin{align}
    \eta_\text{out} = \frac{\kappa_r}{\kappa}\cdot\frac{\kappa}{\kappa+\gamma},\label{eq:strongCouplingEff}
\end{align}
assuming $\kappa_t = 0$ and $\kappa_r$ is the output port of the emitted photons. Equation~(\ref{eq:strongCouplingEff}) separates the contribution of cavity outcoupling and the fractional photon decay through the cavity. Thus, for example, reducing $\kappa_r$ does not help the efficiency as it decreases both factors. \rev{Furthermore, in the strong coupling regime, the wavepacket of photons oscillates with a coupling rate $g$, which makes the indistinguishability significantly susceptible to both $g$ and $\kappa+\gamma$. When designing a cavity for entanglement, optimal parameters should be chosen within the trade-off space, rather than just strengthening the interaction, which often places the system in weak coupling or bad-cavity regimes ($\kappa >> g >> \gamma$)}.

Spin-photon gate and spin-photon projector LBBs, by contrast, are based on the efficient coherent scattering of a single photon or weak coherent state by the emitter, which in turn requires strong interaction between the emitter's dipole and a photonic mode. 
Spatial confinement of an optical mode by means of optical cavities or light-guiding structures (i.e. waveguides or optical fibers) enables this strong interaction. Here we will focus on the use of cavities, as the scattering and emission of an emitter are similar~\cite{kiilerich2019inputoutput,kiilerich2020quantum} except for the numerical calculation of the coupling (see~\cite{sheremet2023waveguide} for multi-emitter cases). In the strong coupling regime, the response of the cavity to incoming photons with the cavity system is modulated in amplitude and phase by the state of the emitter. The cavity-emitter parameters and chosen operation points (in terms of cavity-emitter detuning and photon frequency) can be used together with the system input-output formalism relations (see Appendix and~\cite{kiilerich2019inputoutput,kiilerich2020quantum}) to determine the response of the system, in terms of reflection and transmission coefficients and loss. The spin-dependent reflection and transmission coefficients are complex-valued and describe the amplitude and phase of the photon after the interaction. They are used to realize spin-photon gates and spin-photon projectors. In practice, this can be either changing the amplitude or the phase of the photonic state, conditioned on the spin being in the bright state, as shown in the last two rows of Fig.~\ref{fig:PBB}(a). When modulating phase, it is convenient to use a single-sided cavity as almost all photons will be reflected optimizing the efficiency, while amplitude modulation benefits from a critically coupled cavity maximizing the contrast of the field amplitudes. As anticipated, it is possible that the unwanted transition is close in frequency that it also couples to the cavity. In such cases, the relative phase or amplitude contrast between the system response has to be optimized for high fidelity of the spin-photon emission. This can be done by changing the operation point, i.e. the emitter-cavity detuning and the input photon frequency.

\begin{table*}
\begin{threeparttable}[b]
    \caption{ Physical building blocks as a quantum channel.}
    \label{tab:PBB}
    \centering
    \begin{tabular}{||p{2.57cm}||p{6.44cm}|p{8.54cm}||}
        \hline \hline
         \textbf{PBB} & \textbf{Quantum Channel} & \textbf{Operator} \\
         \hline 
         \multicolumn{3}{|| c ||}{Spin PBB}\\
         \hline
         State Preparation &  $\rho_\text{out} = F_\text{state}\ket{\psi}\bra{\psi} + (1-F_\text{state})\ket{\psi^\perp}\bra{\psi^\perp}$ & $\braket{\psi|\psi^\perp} = 0$ \\
         \hline
         Qubit Error & \multicolumn{2}{l ||}{$\hat{\rho}_\text{out}= F_1\hat{\rho}_\text{in}+\frac{1-F_1}{3}(\hat{\sigma}_x\hat{\rho}_\text{in}\hat{\sigma}_x+\hat{\sigma}_y\hat{\rho}_\text{in}\hat{\sigma}_y+\hat{\sigma}_z\hat{\rho}_\text{in}\hat{\sigma}_z)$}\\
        \hline
        Two-Qubit Error &  \multicolumn{2}{l ||}{$\hat{\rho}_\text{out}= F_2\hat{\rho}_\text{in}+\frac{1-F_2}{15}\sum\limits_{\substack{\hat{A},\hat{B}=\hat{I},\hat{\sigma}_x,\hat{\sigma}_y,\hat{\sigma}_z \\ <\hat{A},\hat{B}>\neq\hat{I}\otimes\hat{I}}}(\hat{A}^\dagger\otimes\hat{B}^\dagger)\hat{\rho}_\text{in}(\hat{A}\otimes\hat{B})$}\\
        \hline
         \multicolumn{3}{|| c ||}{Photonic PBB}\\
                \hline
         Photonic Loss & $\rho_\text{out}= \text{tr}_L\left[\hat{U}(\rho_\text{in}\rev{\otimes\ket{0_L}\bra{0_L}})\hat{U}^\dagger\right]$ & $\hat{U} = \exp[\theta_L(\hat{a}\hat{a}_L^\dagger-\hat{a}^\dagger\hat{a}_L)]$~\tnote{a}\\
         \hline
         Mode Mixing & $\rho_\text{out}=\hat{U}\rho_\text{in}\hat{U}^\dagger$ & $\hat{U} = \exp[\pi/4(\hat{a}\hat{b}^\dagger-\hat{a}^\dagger\hat{b})]$\\
         \hline
         Photodetection & $\rho_\text{out}=\hat{\Pi}\rho_\text{in}\hat{\Pi}^\dagger$ & $\hat{\Pi} = \hat{I}-\ket{0_p}\bra{0_p} ~(\ket{1_p}\bra{1_p})$\\
         \hline
         \rev{Entangled Pair Source (SPDC)} & \rev{$\rho_\text{out} = \hat{S}_2(\zeta)\ket{0000}\bra{0000}\hat{S}^\dagger_2(\zeta)$} & \rev{$\hat{S}_2(\zeta) = \exp\big[\zeta (\hat{a}_H^\dagger \hat{b}_V^\dagger+\hat{a}_V^\dagger \hat{b}_H^\dagger) + \text{h.c.}\big]$}\\
        \hline\multicolumn{3}{|| c ||}{Spin-photon PBB}\\
         \hline
         Spontaneous  & $\rho_\text{out}= C\rho_\text{in}C^\dagger$ & $\hat{C} = \sqrt{p_\text{coh}}\hat{C}_\text{coh} + \sqrt{p_\text{loss}}\hat{C}_\text{loss}$\\
         Emission & \hspace{5mm} + $p_\text{incoh}\hat{C}_\text{incoh}\rho_\text{in}\hat{C}_\text{incoh}^\dagger$ & $\hat{C}_\text{coh}=\ket{0_s}\bra{0_s}\otimes\hat{I}+\ket{1_s}\bra{1_s}\otimes\hat{a}^\dagger$ \\
         (optical $\pi$-pulse)& \hspace{5mm} + $p_\text{2ph}\hat{C}_\text{2ph}\rho_\text{in}\hat{C}_\text{2ph}^\dagger $ & $\hat{C}_\text{loss} = \ket{1_s}\bra{1_s}\otimes\hat{a}_\text{loss}^\dagger$\\
         & & $\hat{C}_\text{incoh}=\ket{1_s}\bra{1_s}\otimes\hat{a}_\text{incoh}^\dagger$\\
         & & $\hat{C}_\text{2ph} = \ket{1_s}\bra{1_s}\otimes\frac{1}{\sqrt{2}}\hat{a}^\dagger \hat{a}^\dagger$\\
         \hline
         Coherent & $\rho_\text{out}= \sum_k P_\beta(k)\hat{C}_\text{incoh}(k)(\hat{C}\rho_\text{in}\hat{C}^\dagger)\hat{C}_\text{incoh}^\dagger(k)\,\tnote{b}\,$& $\hat{C}=\ket{0_s}\bra{0_s}\otimes\hat{I}+\ket{1_s}\bra{1_s}\otimes\hat{D}_a(\alpha)\otimes\hat{D}_{a_\text{loss}}(\alpha_L)$\\
         Scattering & & $\hat{C}_\text{incoh}(k) = \ket{0_s}\bra{0_s}\otimes\hat{I} + \ket{1_s}\bra{1_s}\otimes\frac{1}{\sqrt{k!}}(\hat{a}_\text{incoh}^\dagger)^k$\\
         \hline
         Conditional Phase Reflection & $\rho_\text{out}=\text{tr}_L\left(\hat{C}_z\rho_\text{in}\hat{C}_z^\dagger\right)$ & $\hat{C}_z=\sum\limits_{k_s=0,1}\ket{k_s}\bra{k_s}$\\
         & & \hspace{2 mm} $\otimes\exp(i(\angle r_k \hat{a}^\dagger \hat{a}+\angle l_k \hat{a}_\text{loss}^\dagger \hat{a}_\text{loss}))\exp\left[\theta_k(\hat{a}^\dagger \hat{a}_\text{loss}-\hat{a}\hat{a}_\text{loss}^\dagger)\right] \tnote{c}\,$\\

        \hline \hline

    \end{tabular}
    \begin{tablenotes}
    \item[a] $\theta_L = \arcsin(\sqrt{L})$. \\
    \item[b] $P_\beta(k) = \frac{|\beta|^{2k} \exp^{-|\beta|^2}}{k!}$. Note that this assumes perfect incoherent photon filtering afterward. For the case of imperfect filtering, two incoherent modes for collection and loss should be counted separately (then, filtering modifies the collection incoherent mode only). \\
    \item[c] $\theta_k = \arcsin\left({\sqrt{L_k}}\right)$, \rev{$r_k$ ($l_k$) are the complex coefficient of reflection (loss) of a photon for spin state $\ket{k_s}$. ``$\angle(\cdot)$'' is the phase of the following complex coefficient.}
    \end{tablenotes}
\end{threeparttable}
\end{table*}

\subsection{Quantum channel description of PBB} \label{ssec: quantum channel}
The systematic and modular treatment of quantum channels is the key role of the PBB layer. Quantum channels can be additive and multiplicative: e.g. optical $\pi$-pulse emission spin-photon interface and weak excitation spin-photon interface, respectively. Table~\ref{tab:PBB} lists the representative PBBs, which we use in Sec.~VII for benchmarking REPs. In the following sections, we explain the items of Table~\ref{tab:PBB}.

\subsubsection{Spin PBB}
The state preparation block outputs the desired state $\ket{\psi}$ with fidelity $F_\text{state}$. Our model assumes that a probabilistic error switches the state to an orthogonal state with probability $1-F_\text{state}$. The density matrix of the system is an incoherent mixture of the two components weighted with the probabilities: $\hat{\rho} = F_\text{state}\ket{\psi}\bra{\psi} + (1-F_\text{state})\ket{\psi^\perp}\bra{\psi^\perp}$.

Note that the quantum channel description of PBBs depends on the model of the physical systems. Let's assume that one rotates a perfectly prepared $\ket{0}$ state for the preparation of $\hat{R}_x(\theta)\ket{0}$, where $\hat{R}_x(\theta)$ is the single qubit rotation around the $x$ axis by $\theta$. However, if $\theta$ has a bias error of $\epsilon$, then $\rho_\text{out} = \hat{R}_x(\theta+\epsilon)\ket{0}\bra{0}\hat{R}_x^\dagger(\theta+\epsilon)$ will be more accurate. Choosing a realistic and accurate model is important. Our model-flexible description of PBBs enables the framework to cover all possible quantum hardware, as far as its quantum modeling can be made.

The most widely used error model for qubit operations (gates) is the depolarization channel. This channel maps the portion of qubits' population into a maximally mixed state, which is achieved with the uniformly probable application of Pauli operators (for $n$-qubit gates, the $n$-tensor product of Pauli's and identity operator, except $I^{\otimes n}$). The Table~\ref{tab:PBB} lists one- and two-qubit gate errors as they are often used,
\begin{align*}
\hat{\rho}_\text{out}&= F_1\hat{\rho}_\text{in}+\frac{1-F_1}{3}(\hat{\sigma}_x^\dagger\hat{\rho}_\text{in}\hat{\sigma}_x+\hat{\sigma}_y^\dagger\hat{\rho}_\text{in}\hat{\sigma}_y+\hat{\sigma}_z^\dagger\hat{\rho}_\text{in}\hat{\sigma}_z), \\
\hat{\rho}_\text{out}&= F_2\hat{\rho}_\text{in}+\frac{1-F_2}{15}\sum\limits_{\substack{\hat{A},\hat{B}=\hat{I},\hat{\sigma}_x,\hat{\sigma}_y,\hat{\sigma}_z \\ \hat{A}\otimes\hat{B}\neq\hat{I}\otimes\hat{I}}}(\hat{A}^\dagger\otimes\hat{B}^\dagger)\hat{\rho}_\text{in}(\hat{A}\otimes\hat{B}).
\end{align*}
In the simulation of color centers in Sec. VII, we assume that spin PBBs are perfect because these operations have an order of magnitude smaller errors than photonic or spin-photon PBBs.

\subsubsection{Photonic PBB}
A lost photon while traveling through the fiber by absorption or scattering can project the spin-photon entangled state to a trivial state. A popular way to describe the process is the Lindblad master equations~\cite{scully1999quantum}. This is equivalent to unitary evolution with an auxiliary mode and partial tracing (see the Table element). The unitary evolution is equivalent to the beamsplitter operation, of which the angle is determined from the loss ($\theta_L = \arcsin{\sqrt{L}}$, see Appendix C). The same unitary evolution also describes the mode mixing for erasing `which path information'. Here, we listed the perfect mode mixing with $\theta = \pi/4$, but imperfect mode mixing can be set with different $\theta$ for the biased case, or even combinations of $\theta$s for stochastic description.

We can implement the photodetection on a specific mode ($p$ subscription in the table) with a projection operator. Here, we used the projection of a Fock-state encoded photonic mode, but one can use the eigenstate of a different basis for the projection operator.

\rev{Lastly, we present the PBB for entangled photon-pair sources. Among many implementations and photon encoding, we specifically consider spontaneous parametric down-conversion (SPDC), polarization-encoded photon pair state~\cite{kwiat1995new, couteau2018spontaneous}, 
\begin{align*}
&\rho_\text{out} = \ket{\text{SPDC}}\bra{\text{SPDC}}, \\
&\ket{\text{SPDC}} = \hat{S}_2(\zeta)\ket{0000} = e^{-|\zeta|^2/2} \cdot \Big[\ket{0000} \nonumber \\
&+\zeta\big(\ket{1_{a,H}0_{a,V}0_{b,H}1_{b,V}}+\ket{0_{a,H}1_{a,V}1_{b,H}0_{b,V}}\big)+...\Big],
\end{align*}
where $\hat{S}_2(\zeta) = \exp\big[\zeta (\hat{a}_H^\dagger \hat{b}_V^\dagger+\hat{a}_V^\dagger \hat{b}_H^\dagger) + \text{h.c.}\big]$ is the SPDC operator, $\zeta$ denotes the numerical parameter determined by pump field, $\hat{a}$ and $\hat{b}$ are two spatial photonic modes, $H$ and $V$ indicate two polarization modes, and ``h.c.'' refers to Hermitian conjugate terms. It is worth noting that $\ket{HV}+\ket{VH} = \ket{1_{a,H}0_{a,V}0_{b,H}1_{b,V}}+\ket{0_{a,H}1_{a,V}1_{b,H}0_{b,V}}$ in the literature for simplicity, whereas we explicitly specify photon numbers in each photonic mode to align with other PBBs.
}

\subsubsection{Spin-photon PBB}
The ideal operation of emission spin-photon interface using optical $\pi$-pulse is described with the coherent channel subscripted with ``coh''. When the spin is in the dark state ($\ket{0}$) it does not affect the state of the photonic mode ($\hat{I}$). When the spin is in the bright state ($\ket{1}$), the desired outcome is the creation of a single photon in the mode of interest, $\hat{a}$. This results in the Kraus operator $\hat{C}_\text{coh}=\ket{0_s}\bra{0_s}\otimes\hat{I}+\ket{1_s}\bra{1_s}\otimes\hat{a}^\dagger$.

In realistic devices, ideal events only happen with probability $p_\text{coh}$. With probability, $p_\text{incoh}$, the incoherent channel, labeled with ``incoh'', adds a photon with a random phase that does not coherently interfere, and it reflects the non-unity indistinguishability. This can be modeled with the addition of photon to the incoherent mode $\hat{a}_\text{incoh}$ that will not be interfered with by the mode mixing PBB. Likewise, double excitation and finite efficiencies additively contribute to the density matrix through $\hat{C}_\text{2ph}$ and $\hat{C}_\text{loss}$, with $p_\text{2ph}$ and $p_\text{loss}$ (see Table~\ref{tab:PBB} and Appendix D1). Again, these PBBs can add more channels such as ionization, or remove some if not wanted. 

All the parameters including probabilities are calculated at the quantum optical modeling layer, and PBBs require these numerical values for the calculation of $\rho_\text{out}$. See Appendix~\ref{Apdx:QOParameter} for the calculation of PBB parameters from physically characterized ones.

Instead of a short strong optical pulse, one can weakly drive the transition by coherently scattering the field. The scattered field by this way is in a coherent state with amplitude $\alpha$. The process can be encapsulated by the displacement operator $\hat{D}_{\hat{a}}(\alpha)$ where $\hat{a}$ represents the mode that the operator acts on ($\hat{D}(\alpha)\ket{0} = \ket{\alpha}$). At the same time, the excitation field is also scattered to unwanted mode $\hat{a}_\text{loss}$. Moreover, incoherent scattering due to noises adds photons to another mode with Poisson statistics. We can express the loss-mode part of the density matrix as,
\begin{align*}
\hat{\rho}_\text{loss} = \sum_i \frac{n_\text{loss}^i \exp{(-n_\text{loss})}}{i!}\ket{i}\bra{i},
\end{align*}
\rev{where $n_\text{loss}$ is the average number of photons added to the loss mode. $\ket{n\neq 0}$ components of the density matrix reduce the fidelity of the fidelity of spin-photon entanglement and subsequent spin-spin entanglement, because it leaks the state information (only bright state, $\ket{1_s}$, adds the photon to the loss mode).}

\section{Software implementation of the framework}

The layered and modular framework lends itself well to implementation in simulation code. Our QuREBB (Quantum Remote Entanglement Building Blocks) simulation repository is available on GitHub~\cite{beukers2023qurebb}, and we used it to simulate and compare different REPs (Sec.~VII). One can reuse implemented codes with minimal corrections due to our modular framework, even for different protocol topologies (with modifications of LBBs) or different physical systems (with modifications of quantum modeling).

The code is based on the QuTiP  package~\cite{johansson2012qutip} written in Python. For the simulation of REPs, it is important to represent composite quantum systems (involving stationary qubits and multiple photonic modes), track all modes, and trace out photonic loss modes when needed. QuTiP references subsystems by indices and partial traces modify the indices. We improved the indexing capabilities of the quantum object (Qobj) in QuTiP and refer to a quantum mode with a string such that each subsystem can be properly named (dictionary). We created an inherited class ``named quantum object'' (NQobj) compatible with QuTiP standard functions and described the names of the subsystems in an attribute 'names'. With this new attribute in place, one can allow for operations with NQobjs of different sizes. For example, when a density matrix evolves with a unitary, one can omit the identities for the subsystems not involved as this is inferred from the names of the objects. For a modular use of the code, this is an essential feature as this allows one to write code for the building blocks without knowledge of the whole system. Moreover, it allows to trace out loss modes anywhere in the calculations as they can be indexed by name, without the risk of shifting all the indices of the other modes by removing the loss mode.

The state of the system is represented by a non-normalized density matrix.  In this way both the state and the success probability are represented by one object. The state is the normalized density matrix and the success probability is the trace of the density matrix. These non-normalized density matrices act as the interface between the physical building blocks, which are the quantum operators on the density matrix.

The quantum modeling of the spin-photon interface favors the description with creation and annihilation operators (Heisenberg picture), while the non-unitary evolution of quantum systems favors density matrices (Schr\"{o}dinger picture). To bridge the gap one needs descriptions like one of the beamsplitters in Table~\ref{tab:PBB}: $\hat{U} = \exp[\theta(\hat{a}\hat{b}^\dagger-\hat{a}^\dagger\hat{b})]$. We modeled the cavities as a composite system of beamsplitters and phase shifters (see Appendix~\ref{apdx:SPIUnitaries}). The method is advantageous in directly describing arbitrary big Fock-state space in QuTiP as users can choose the size of the creation and annihilation operators accordingly. The code is directly extendable to the simulation of weak coherent states with non-negligible multi-photon states.

While our package covers the low-level description of quantum systems to the protocol-level description of elementary entanglement links, it complements other tools at the link layer or above. This includes Netsquid~\cite{coopmans2021netsquid}, an event-driven network simulator that can cover link layer~\cite{dahlberg2019_link_layer} or higher level with a large number of nodes~\cite{wehner2018_quantum_internet}, as well as QuNetSim~\cite{diadamo2021qunetsim} and the others~\rev{\cite{matsuo2019quantum, bartlett2018distributed}}.

\section{Simulating and Benchmarking Entanglement Protocols}
\label{sec:simulating}

In this section, we show an example of how the proposed framework can be used to break down three different entanglement protocols into realistic physical building blocks, and we use the software introduced in the previous section to simulate the performance and compare them. We set the physical platform used for all the protocols to be the silicon-vacancy center in a diamond coupled to an optical cavity.

\begin{figure*}
    \centering
    \includegraphics[width=.98\linewidth]{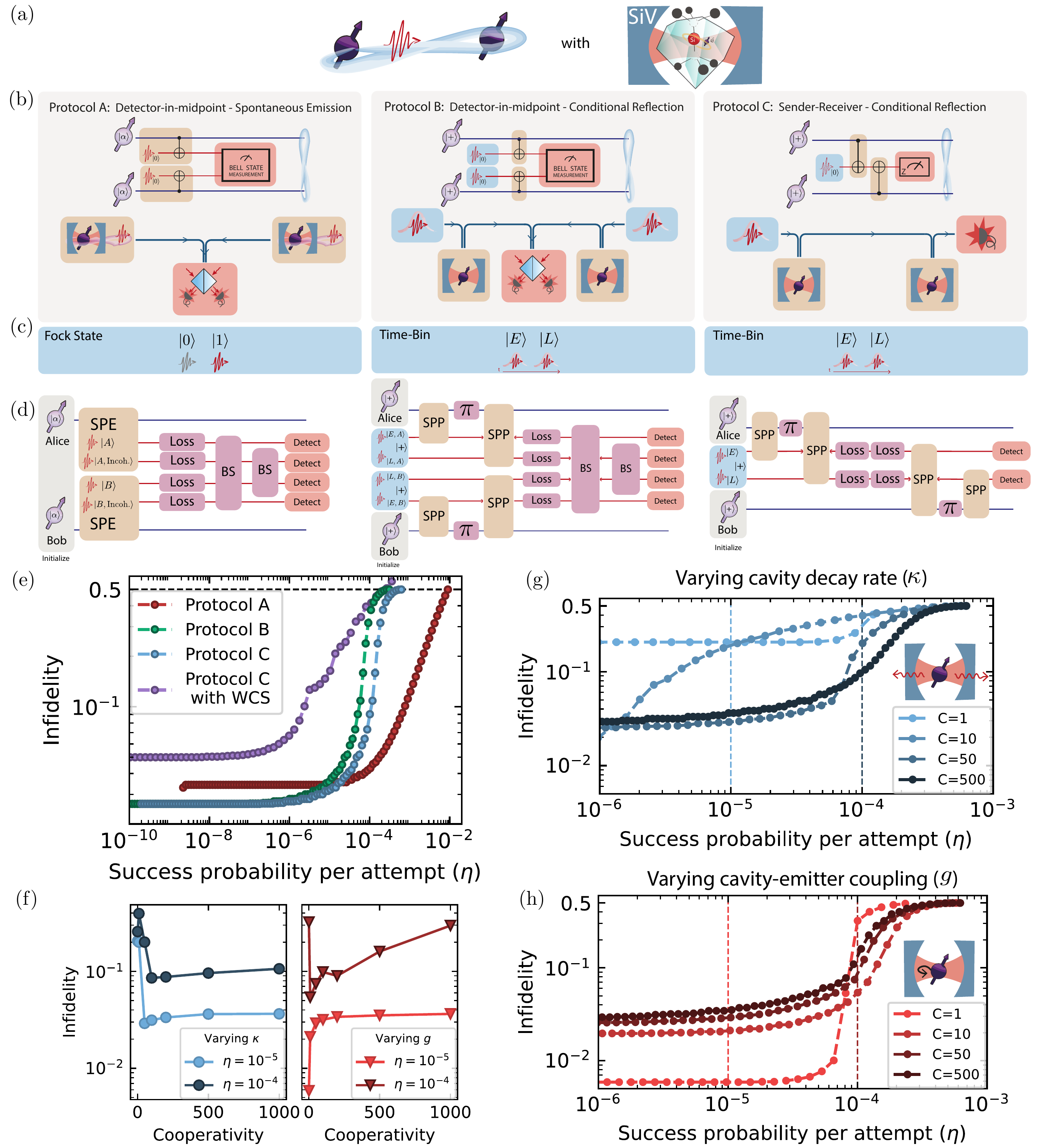}
    \caption{ Comparison of different protocols for entanglement generation using the SiV in diamond coupled to an optical cavity as spin-photon interface. (b) Circuit diagram and logical building block descriptions of the protocols under investigation. (c) Photonic encodings for each protocol. (d) The protocol after the LBBs are compiled to hardware-aware PBBs and includes imperfections such as loss.
    \rev{The photon states are labeled by node of interaction ($A$ and $B$ for Alice and Bob) by time-bin encoding ($E$ and $L$ for Early and Late mode), and whether the photon is resulting from incoherent emission ($\textrm{Incoh.}$). The initial photon state is defined as $|+\rangle = 1/\sqrt{2}(|E\rangle + |L\rangle)$.} (e) Simulated \rev{success probability}-infidelity curves for the different protocols using the parameters of Table~\ref{tab:notations}. \rev{Protocol C is also simulated for a weak coherent state (WCS) input instead of single photons.} (f, g, h) Performance of Protocol C when the cooperativity is changed by either improving the cavity decay rate $\kappa$, or the emitter-cavity coupling $g$, following Eq.~(\ref{eq:cooperativity}). \rev{(f) Shows the fidelity for two different success probabilities $\eta$ when varying $\kappa$ or $g$ (left and right graph, respectively) (g) and (h) report the success probability-infidelity curves for some of the cooperativities. The dashed lines mark the values of $\eta$ shown in (f).} }
    \label{fig:Comparison1}
\end{figure*}

We outline our simulation comparison in Fig.~\ref{fig:Comparison1}. The hardware platform we chose is the SiV color center in diamond, and we consider photonic crystal cavities to either enhance the spin-photon emission or to realize spin-photon projectors in the strong coupling regime. The protocols we compare are one emission-based protocol with detection-in-midpoint topology, where we use Fock state encoding for the photon (Protocol A in Fig.~\ref{fig:Comparison1}(b) and two projector-based (Protocols B and C) with the conditional amplitude reflection and the time-bin encoding, with a sender-receiver and detection-in-midpoint topology. The detailed implementation in terms of physical building blocks is shown for all protocols (Figures~\ref{fig:Comparison1}(c) and (d)). \rev{More details on the quantum optical modeling used can be found in Appendices \ref{apdx:SPIUnitaries} and \ref{Apdx:QOParameter}.}

We compare the performance of the protocols according to the simulation results. For each simulation, we start with realistic experimental values for this system that can be found in the literature (see Table~\ref{tab:notations}), and we search the optimal operation point of the spin-photon interface by sweeping a subset of the parameters\rev{("sweep parameters", indicated as \textit{variable} in Table~\ref{tab:notations}),} and for each configuration calculate the protocol outcome. We choose to optimize only parameters that are easily controllable in an experimental setting: the detuning between cavity and emitter and the detuning between input photon and cavity (which we call the operation frequency). The frequency difference between the two optical transitions involved, $\delta_{01}$, can in principle be tuned by applying an external magnetic field and this can have a big impact on the protocol performance. However, as the magnetic field is intertwined with other features of the system, such as the cyclicity of the optical transition and the qubit frequency, we do not consider it here as \rev{a sweep parameter}. \rev{Similarly, intrinsic properties of the cavity or the emitter are considered fixed parameters for the protocol optimization.} Ultimately, the simulation gives us the infidelity and success probability with which the protocol achieves a definite target state, which is one of the four Bell states, as a function of the swept parameters. To benchmark the protocols we compare the trade-off between infidelity and success probability. \rev{We report the success probability of the protocol for each attempt, as obtained from the density matrix of the final state. The} effective rate can be extracted from the success probability by accounting for the repetition rate of the protocol, this can give a more consistent comparison.
Figure~\ref{fig:Comparison1}(e) shows the rate-infidelity curves for the three protocols with the starting values reported in Table~\ref{tab:notations}. 
From these curves, one can see how, with these parameters and protocol implementations, Protocol A is advantageous with respect to the others\rev{, except at low rates where the fidelity is limited by the incoherent emission due to dephasing. Here Protocols B and C allow to obtain lower infidelity.}
We also simulate Protocol C using a weak coherent state \rev{(WCS)} input to approximate single photons. In this case, the success probability \rev{drops faster than in the single photon implementation, as it scales with the average photon number in the WCS, and the infidelity is higher since the multiphoton components in the input cause protocol errors}.

Finally, we simulate how Protocol C performs for different values of the cooperativity. Again, we optimize for each point by sweeping the \rev{``sweep parameters"} as above \rev{and, in addition to that, we vary only one of the intrinsic cavity parameters to change the cooperativity}. The rate-infidelity curves are reported in Fig.~\ref{fig:Comparison1}(g) and (h). We test two ways to change the cooperativity, as can be seen from Eq.~(\ref{eq:cooperativity}). In (g) we do this by varying the \rev{cavity decay} rate $\kappa$, which is related to the cavity quality factor $Q$, at fixed cavity-emitter coupling $g$. In (h) we keep $\kappa$ fixed and we vary the cavity-emitter coupling $g$, which can practically be realized by changing the cavity mode volume, overlap between the emitter dipole and the field distribution, or the quantum efficiency and Debye-Waller factor of the optical system. In Fig.~\ref{fig:Comparison1} (f) we report the infidelity at two different \rev{success probabilities}. Interestingly, \rev{increasing the cooperativity by only optimizing a subset of the parameters does} not always lead to better performance. This can be attributed to the Purcell broadening of the optical linewidth which, when a second optical transition is close in frequency ($\delta_{01}$ is small compared to the Purcell broadened linewidth \rev{and cannot be optimized}), can cause a decrease in fidelity due to undesired interaction with the wrong spin state. Varying the cavity linewidth ($\kappa$) or varying the coupling $g$ can affect this phenomenon differently, especially at low cooperativities. \rev{Furthermore, when increasing the cooperativity in Fig.~\ref{fig:Comparison1} by only changing either $\kappa$ or $g$, we pass from the bad cavity regime ($\kappa \gg g$) to the strong coupling one ($g \gg \kappa$). As discussed in Sec. \ref{subsec:cavity}, this regime is not necessarily optimal for entanglement generation, especially if the system is not optimized accordingly.}

\section{Conclusions}
In this work, we introduced a modular framework to describe photon-mediated remote entanglement protocols. Our framework divides the remote entanglement protocols into four different layers, allowing a perspective of the entanglement protocol from an abstract and hardware-agnostic overview to a detailed description of each component and physical implementation. We describe the function of the different layers and provide examples of their realization, including detailed modeling of cavity-based spin-photon interfaces. Finally, using a software implementation that directly reflects the modular approach of the framework, we simulated different remote entanglement protocol topologies based on a realistic experimental platform and investigated how they perform under different parameter regimes.

For future work, our framework can expand into both higher-level uses and lower-level descriptions. At the higher level, leveraging the logical operations of the LBB layer can construct error-corrected encoded quantum networks~\cite{jiang2009quantum}. Incorporating error correction schemes into our framework would enable \rev{the simulation of} fault-tolerant quantum applications. On the other hand, the integration of material-level details of physics, for example, \cite{thiering2018initio} for the case of color centers, into our quantum optical modeling can help the choice of quantum systems to optimally configure the network. This inclusion of finer-grained aspects will enhance the accuracy of our simulations as well. \rev{In addition, we believe our framework can be extended to apply to continuous variable entanglement distribution~\cite{dias2020quantum} and qudit entanglement distribution~\cite{zheng2022entanglement}.}
\\

\section{Acknowledgments}
We thank M. Ruf , H. Sharma, W. Dai and A. S. Sørensen for the fruitful discussions and H. Bernien, T. E. Northup,  S. Krastanov, K.C. Chen and I.B. Harris for critically reviewing the manuscript.

We acknowledge funding from the Dutch Research Council (NWO) through the project “QuTech Part II Applied-oriented research” (project number 601.QT.001), the Spinoza prize 2019 (project number SPI 63-264) and the Zwaartekracht program Quantum Software Consortium (project no. 024.003.037/3368). We also acknowledge funding from the Dutch Ministry of Economic Affairs and Climate Policy (EZK), as part of the Quantum Delta NL programme, and Holland High Tech through the TKI HTSM (20.0052 PPS) funds, and from the joint research program “Modular quantum computers" by Fujitsu Limited and Delft University of Technology, co-funded by the Netherlands Enterprise Agency under project number PPS2007.

This work was partially supported by AFOSR grant FA9550-20-1-0105 supervised by Gernot Pomrenke, the NSF Center for Quantum Networks awarded under cooperative agreement number 1941583, and Cisco Research. H.C. acknowledges the Claude E. Shannon Fellowship and the Samsung Scholarship. D.E. also acknowledges funding from the NSF C-ACCEL program.

\bibliography{main}
\bibliographystyle{apsrev4-2}

\newpage

\onecolumngrid
\appendix
\section{Nomenclature}

\begin{table}[H]
    \centering
    \caption{Nomenclature used in this tutorial}
    \begin{tabular}{|l|l|}
    \hline
        \multicolumn{2}{| c |}{\textbf{Topology}}\\ \hline
        \multicolumn{2}{| c |}{Detection-in-midpoint} \\ \hline
        \multicolumn{2}{| c |}{Sender-receiver} \\ \hline
        \multicolumn{2}{| c |}{Source-in-midpoint} \\ \hline
        \multicolumn{2}{| c |}{\textbf{Logical Building Blocks}} \\ \hline
        Spin & Initialization \\ \hline
        ~ & Gate \\ \hline
        ~ & Measurement \\ \hline
        Photon & Photon source \\ \hline
        ~ & Photon pair source \\ \hline
        ~ & Measurement \\ \hline
        ~ & Bell state measurement \\ \hline
        ~ & Photon gate \\ \hline
        Spin-Photon & Spin-photon emission \\ \hline
        ~ & Spin-photon gate \\ \hline
        ~ & Spin-photon projector \\ \hline
        ~ & Spin-photon absorption \\ \hline
        \multicolumn{2}{| c |}{\textbf{Encoding}} \\ \hline
        \multicolumn{2}{| c |}{Fock state} \\ \hline
        \multicolumn{2}{| c |}{Time-bin}  \\ \hline
        \multicolumn{2}{| c |}{Polarization}  \\ \hline
        \multicolumn{2}{| c |}{Dual-rail}  \\ \hline
        \multicolumn{2}{| c |}{Frequency}  \\ \hline
        \multicolumn{2}{| c |}{\textbf{Physical Building Blocks}}  \\ \hline
        Spin & Initialization \\ \hline
        ~ & Gate \\ \hline
        ~ & Measurement \\ \hline
        Photon  & Initialization \\ \hline
        ~ & Measurement \\ \hline
        ~ & Mode mixing \\ \hline
        ~ & Z-Rotation \\ \hline
        ~ & X-Rotation \\ \hline
        Spin-photon & Spontaneous emission \\ \hline
        ~ & Coherent scattering \\ \hline
        ~ & Raman scattering \\ \hline
        ~ & Conditional amplitude reflection \\ \hline
        ~ & Conditional phase reflection \\ \hline
    \end{tabular}
\end{table}
\FloatBarrier
\newpage
\section{Notation and simulation parameters }

\begin{table}[h!]
\begin{threeparttable}[b]
\caption{SiV parameters for simulations}
\label{tab:notations}

\centering

\begin{tabular}{||p{5cm}||p{2cm}|c c||}
\hline \hline
\textbf{Name} & \textbf{Symbol} & \textbf{Spin-Photon Projector}  & \textbf{Spin-Photon Emission}\\
\hline
Emitter Resonance Frequency & $\omega_a$ & \multicolumn{2}{c ||}{406.706 THz} \\
Cavity Resonance Frequency & $\omega_c$ & \multicolumn{2}{c ||}{Variable} \\
Laser Frequency & $\nu$ & \multicolumn{2}{c ||}{Variable} \\
Laser-Cavity Detuning & $\Delta_{lc}$ & \multicolumn{2}{c ||}{Variable} \\
Cavity-Emitter Detuning & $\Delta_{ce}$ & \multicolumn{2}{c ||}{Variable} \\
Laser-Emitter Detuning & $\Delta_{le}$ & \multicolumn{2}{c ||}{Variable} \\
Emitter Radiative Decay Rate (ZPL)& $\gamma_r$ & \multicolumn{2}{c ||}{13.1 MHz~\tnote{a}}\\
Emitter Total Decay Rate & $\gamma$ & 92.5 MHz~\tnote{a} & 100 MHz\\
Pure Dephasing & $\gamma^*$ & \multicolumn{2}{c ||}{30.5 MHz~\tnote{b,c}}\\
Spectral Diffusion & $\sigma_\omega$ & \multicolumn{2}{c ||}{N/A\tnote{d}}\\
Total Emitter Linewidth & $\Gamma$ & 123 MHz \tnote{b} & N/A\\
Optical Transitions Detuning & $\delta_{01}$ & \multicolumn{2}{c ||}{1 GHz}\\
Emitter-Cavity Coupling & $g$ & 8.38 GHz \tnote{b} & 6.81 GHz\\
Cavity Decay Rate & $\kappa$ & 21.8 GHz \tnote{b} & 
 $\kappa_l+\kappa_t = 89$ GHz, $\kappa_r = 240$ GHz \tnote{e} ~\\
Cavity Quality Factor & $Q$ & 18,700 \tnote{b} & 1,237 \\
Cavity Cooperativity & $C$ & 105 & 4.3 \tnote{f}\\
Debye-Waller Factor & DW & \multicolumn{2}{c ||}{0.7 \tnote{g}}\\
Quantum Efficiency & QE & \multicolumn{2}{c ||}{0.2} \\
Link Loss & - & \multicolumn{2}{c ||}{0.9}\\
Device Insertion Loss & - & \multicolumn{2}{c ||}{0.5}\\
\hline \hline
\end{tabular}

\begin{tablenotes}
\item [a] \cite{thiering2018initio} \\
\item [b] \cite{bhaskar2020_repeater} \\
\item [c] This data for over-coupled cavity does not exist. We assume that it is the same as the critically-coupled case.\\
\item [d] To the best of our knowledge, there are no spectroscopic results resolving pure dephasing and spectral diffusion of SiV. In this work, we simply assume that the broadening in \cite{bhaskar2020_repeater} is from the pure dephasing considering the Lorentzian lineshape. \\
\item [e] \cite{knall2022efficient} \\
\item [f] Note that this number is converted with pure dephasing assumed. \\
\item [g] \cite{evans2018integrated}
\end{tablenotes}
\end{threeparttable}
\end{table}
\FloatBarrier

\section{Modeling conditional reflection spin-photon interface with unitaries}
\label{apdx:SPIUnitaries}

\begin{figure}
    \centering
    \includegraphics[width=0.35\textwidth]{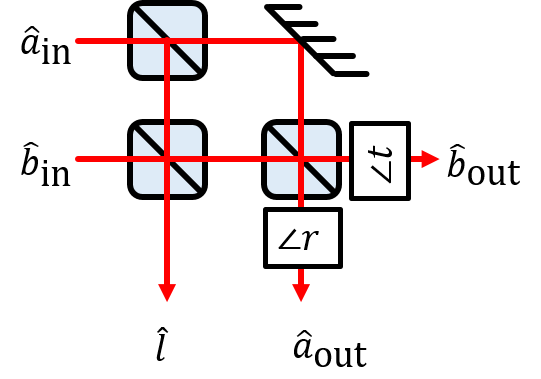}
    \caption{Two-port-to-three-port unitary construction with beamsplitters.}
    \label{fig:2to3}
\end{figure}

\rev{We model the conditional reflection spin-photon interfaces using input-output formalism~\cite{kiilerich2019inputoutput,kiilerich2020quantum}. This describes the response of the system in terms of$r$, $t$, and $l$ as the complex coefficients of reflection, transmission, and loss, respectively. To calculate the unitary operator in the Fock state basis, we use the beamsplitter configuration in Fig.~\ref{fig:2to3} and the description of a beamsplitter
\begin{equation}
 U = \exp{\left[\theta(a^\dagger b- a \rev{b}^\dagger)\right]}   
\end{equation}
} $L=|l|^2$ is the loss. For example, if 30\% of photons are lost, then $L=0.3$.
\begin{align}
    U_1 = \exp{\left[\theta_1(a^\dagger l- a \rev{l}^\dagger)\right]}. \\
    \theta_1 = \arctan\left(\frac{\sqrt{L}}{\sqrt{1-L}}\right).
\end{align}
Then, from $r$ and $t$, we calculate the normalized $r'$ and $t'$;
\begin{align}
    r' = r/\sqrt{|r|^2+|t|^2} \\
    t' = t/\sqrt(|r|^2+|t|^2).
\end{align}
The unitary for the splitting is
\begin{align}
    U_2 = \exp{\left[\theta_2(a^\dagger b- ab^\dagger)\right]} \\
    \theta_2 = \arctan\left(\frac{|t'|}{|r'|}\right)
\end{align}
For adjusting the phases,
\begin{align}
    U_3 = \exp\left[i(\angle r a^\dagger a +\angle t b^\dagger b)\right]
\end{align}
The total unitary for loss-reflection-transmission is,
\begin{align}
    U = U_3\cdot U_2\cdot U_1.
\end{align}

One can use both ports of the cavity with dual rail encoding. The interference of fields is reflected in the interference of the reflection/transmission coefficient derived from the coherence of the spins. Figure~\ref{fig:2to3} shows the two-port-to-three-port unitary decomposition. The splitting ratio of BS1 is determined by the loss of input port 1, while that of BS2 is by the loss of input port 2. BS3 combines the output ports of BS1 and BS2 to generate reflection and transmission. A total of five phase shifters are added between the beamsplitters and at the output port for the phase adjustment. 

For a larger number of input/output ports, we can use Reck~\cite{reck1994experimental} or Clements~\cite{clements2016optimal} decomposition. 

\section{Quantum Modeled Physical Parameters} \label{Apdx:QOParameter}
Here, we discuss the PBB parameters calculated from the quantum modeling layer.

\subsection{Spontaneous emission PBB}
\begin{align}
p_\text{coh} &= \bra{1}\rho_\text{in}\ket{1}\cdot\frac{\kappa_r}{\kappa}\cdot\frac{C}{C+1}\cdot \frac{F_p\gamma_r}{\Gamma+F_p\gamma_r} + \bra{0}\rho_\text{in}\ket{0} \\
p_\text{incoh} &= \bra{1}\rho_\text{in}\ket{1}\cdot\frac{\kappa_r}{\kappa}\cdot\frac{C}{C+1}\cdot\frac{\gamma*+\sigma_\omega}{\Gamma+F_p\gamma_r}\\
p_\text{2ph} &= 0 \\
p_\text{loss} &= 1-p_\text{coh}-p_\text{incoh}-p_\text{2ph}\\
\end{align}
The double excitation probability $p_\text{2ph}$ is a function of the lifetime, pulse-width, and ionization probability. Here, we neglect the two-photon emission for simplicity. This is possible by adjusting the pulse area so that one transition is odd multiple $\pi$-pulse and the other is even multiple.

\subsection{Coherent scattering PBB}
\begin{align}
\alpha_\text{tot} &= \alpha+\alpha_L \\
\alpha_\text{tot} &= \eta\alpha \\
\alpha_L &= \frac{1-\eta}{\eta}\alpha \\
\eta &= \frac{\kappa_r}{\kappa}\cdot\frac{F_p}{F_p+1} \\
|\beta|^2 &= \left[\frac{\gamma*}{\Gamma'-\gamma*}+\frac{1-QE'\cdot DW'}{QE'\cdot DW'}\right]|\alpha_\text{tot}|^2 \label{eq:alphaToBeta}\\
\Gamma' &= \Gamma + F_p\gamma_r\\
DW' &= \frac{(F_p+1)DW}{F_p\cdot DW + 1}\\
QE' &= \frac{QE\cdot(F_p\cdot DW+1)}{1+F_p \cdot QE\cdot DW}
\end{align}
$\eta$ is the collection efficiency of the coherent emission. Prime represents the cavity-modified rate or ratio of transitions. In Eq.~(\ref{eq:alphaToBeta}), the first term is for incoherent emission, and the second term is for inefficiencies. 

\rev{
\section{Simulations parameter sweep and details}

To optimize the fidelity and success probability of the simulations in Fig.~\ref{fig:Comparison1} we perform the parameters sweep in Table~\ref{tab:sweepparams}.

\begin{table}[h!]
    \centering
    \begin{tabular}{|c|c|c|c|}
    \hline
         Protocol & Parameters & Range & Number of points \\
         \hline
         A & $\alpha$ (initial spin state) & [$10^{-7}, 0.3$] & 500\\
         \hline
         B,C & $\Delta_{la}$ & $[-18, 0]$ GHz & 1000\\
         & $\Delta_{ac}$ & $[0, 120]$ GHz & 60 \\
         \hline
         C (WCS) & $\alpha$ (WCS) & $[0.001, 2]$ & 10\\
         & $\Delta_{la}$ & $[-15, -2]$ GHz & 1200\\
         & $\Delta_{ac}$ & $[0, 120]$ GHz & 50 \\
         \hline
    \end{tabular}
    \caption{Sweep parameters for simulations of Fig.~\ref{fig:Comparison1}}
    \label{tab:sweepparams}
\end{table}

For the cooperativity sweep by varying $\kappa$ we use the same values as in the table. For the sweep by varying $g$ the cavity-emitter spectrum changes significantly, therefore the range has to be optimized for each simulation. The ranges can be found in the simulation notebooks~\cite{beukers2023qurebb}.\\

The simulation were ran on a desktop computer (Intel Xenon CPU, 3.50 GHz, Quad-core, 8-threads, 32 GB RAM). The simulation of a single protocol run takes $\sim350$ ms for Protocols A and C and $\sim550$ ms for Protocol B (mostly dependent on the size of the Hilbert space). By using basic Python multiprocessing functions we can speed up the parameter sweep by a factor $\sim5$ compared to just looping the protocol over the whole parameter space. This results on a runtime of $\sim2$ min for Protocol A, $\sim2$ hours for Protocol B, $\sim1$ hour for protocol C and $\sim10$ hours for the WCS simulation.
}
\end{document}